\documentclass{article} 

\usepackage{amsmath,graphicx,psfrag} 
\usepackage{amssymb}

\begin{document}

\begin{titlepage}

\begin{flushright}
BUTP/2000-16\\
\end{flushright}
\vspace{1cm}

\begin{center}
{\LARGE Fixed Point Gauge Actions with Fat Links:\\ 
Scaling and Glueballs}
\footnote{Work supported in part by Schweizerischer Nationalfonds}
  
\vspace{1cm}
{\large
Ferenc Niedermayer\footnote{On leave from the Institute of Theoretical
Physics, E\"otv\"os University, Budapest},
Philipp R\"ufenacht and
Urs Wenger
}
\\
\vspace{0.5cm}
Institute for Theoretical Physics \\
University of Bern \\
Sidlerstrasse 5, CH-3012 Bern, Switzerland

\vspace{0.5cm}
\vspace*{0.5cm}

{\bf Abstract}
\end{center}
\vspace{-5mm}

\begin{quote}
  A new parametrization is introduced for the fixed point (FP) action
  in SU(3) gauge theory using fat links.
  We investigate its scaling properties by means of the static 
  quark-antiquark potential and the dimensionless quantities 
  $r_0 T_c$, $T_c/\sqrt{\sigma}$ and $r_0 \sqrt{\sigma}$,
  where $T_c$ is the critical temperature of the deconfining phase transition,
  $r_0$ is the hadronic scale and $\sigma$ is the effective string
  tension. These quantities scale even on lattices as coarse as 
  $a \approx 0.3$ fm. 
  We also measure the glueball spectrum and obtain $m_{0^{++}}=1627(83)$ MeV
  and $m_{2^{++}}=2354(95)$ MeV for the masses of the scalar and tensor 
  glueballs, respectively.
\end{quote}
 
\vfill
\end{titlepage}

\section{Introduction}

One way to study quantum field theories beyond perturbation theory is
to discretize the Euclidean space-time, using the lattice spacing $a$
as an ultraviolet regulator \cite{Wilson:1974sk}.
Accordingly, the continuum action is replaced by some discretized
lattice action. The basic assumption of universality means
that the physical predictions -- obtained in the continuum limit 
$a\to 0$ -- do not depend on the infinite variety of discretizing
the action. At any finite lattice spacing, however, the discretization
introduces lattice artifacts. On dimensional grounds one expects
that in purely bosonic theories these discretization errors
go away as ${\cal O}(a^2)$, while in theories with fermions as ${\cal O}(a)$.

Sometimes it is stated that the lattice artifacts in pure Yang-Mills
theories can be beaten simply by brute force -- using the standard
Wilson gauge action and a sufficiently small lattice spacing,
i.e.~by large computer power and memory.
This is only partially true -- for example when one calculates the pressure
of a hot gluon plasma the computer cost grows like $1/a^{10}$
therefore the size of lattice artifacts becomes crucial.

Naturally, one can use the freedom in discretizing the action to minimize the
artifacts. It has been shown by Symanzik
\cite{Symanzik:1983dc,Symanzik:1983gh} that the leading lattice artifacts can
be cancelled in all orders of perturbation theory by tuning the coefficients
of a few dimension $d+2$ operators in bosonic theories (or dimension $d+1$
operators for fermions).  This improvement program can be extended to the
non-perturbative regime \cite{Luscher:1997ug,Luscher:1998pe}.

A different approach, based on renormalization group (RG) ideas
\cite{Wilson:1974jj}, has been suggested in ref.~\cite{Hasenfratz:1994sp}.  By
solving the fixed point (FP) equations for asymptotically free theories one
obtains a classically perfect action -- i.e.~which has no lattice artifacts on
the solutions of the lattice equations of motion.  (One can say that the FP
action is an on-shell tree-level Symanzik improved action to all orders in
$a$.)  Although it is not {\em quantum} perfect, one expects the FP action to
perform better in Monte Carlo simulations.  This is indeed true in all cases
investigated.

The FP approach has been successfully applied to the two-dimensional
non-linear $\sigma$-model \cite{Hasenfratz:1994sp,Blatter:1996ik} and the
two-dimensional CP$^3$-model \cite{Burkhalter:1996dr}. For SU(3) gauge theory
the classically perfect FP action has been constructed and tested in
\cite{DeGrand:1995ji,DeGrand:1995jk,DeGrand:1996ab,Blatter:1996ti} and the
ansatz has been extended to include FP actions for fermions as well
\cite{Bietenholz:1996cy,Bietenholz:1996qc,DeGrand:1997nc}. In the case of
SU(2) gauge theory the FP action has been constructed in
\cite{DeGrand:1996ih,DeGrand:1996zb,DeGrand:1997gu} and its classical
properties have been tested on classical instanton solutions, both in SU(2)
and SU(3) \cite{Farchioni_Papa:98}.

The FP action is not unique -- it depends on the RG transformation chosen, and
it is crucial to optimize the RG transformation to obtain
an interaction range of the action as small as possible.
The value of the FP action on a given field
configuration can be calculated precisely by a classical saddle point
equation.  However, this step is too slow to embed into a Monte Carlo
calculation and one has to invent a sufficiently fast but at the same time
accurate enough method to calculate the FP action.

In earlier works on SU(3) gauge theory the parametrization of the FP action
used Wilson loops and their powers.  We investigate here a new, richer and
more flexible parametrization using plaquettes of the original and of smeared
(``fat'') links.  This describes the FP action more accurately than the loop
ansatz.  For the RG transformation we choose the one investigated in
ref.~\cite{Blatter:1996ti} since the blocking kernel used there (and the
resulting FP action) has better properties than the ``standard'' Swendsen
blocking which uses long staples.  Here we approximate the same FP action with
the new parametrization.  We also improve the method of fixing the parameters:
besides the known action values (for a given set of configurations) we also
fit the known derivatives $\delta {\cal A}^{\text{FP}}(V) / \delta V_\mu(n)$,
i.e.~we have a much larger set of constraints than previously.

Although being much faster than the loop parametrization of comparable
richness, this parametrization has a significant overhead compared to the
Wilson action.  Therefore it is not clear whether it is not better to use in
pure gauge theory a faster but less accurate parametrization of the FP action.
However, in QCD the cost associated with fermionic degrees of freedom
dominates and one can afford a relatively expensive gauge action. In addition,
with fermions it is much harder to decrease the lattice spacing, hence it
could pay off to have a better gauge action as well. (Of course, because of
the ${\cal O}(a)$ artifacts, it is even more important to improve the
fermionic part.)

The paper is organized as follows. In section 2 we present the construction
and parametrization of the FP action.  Section 3 deals with the measurement of
the critical couplings $\beta_c$ of the deconfining phase transition
corresponding to temporal extensions $N_\tau=2$, $3$ and $4$, at various
spatial volumes.  In section 4 we measure the static $q\bar{q}$ potential by
using a correlation matrix between different (spatially) smeared gauge
strings.  In section 5 the scaling properties of the dimensionless quantities
$r_0 T_c$, $T_c/\sqrt{\sigma}$ and $r_0 \sqrt{\sigma}$ are presented.  In
section 6 the low lying glueball spectrum is measured in all symmetry
channels.  For the Wilson action the lowest lying $0^{++}$ state shows
particularly large cut-off effects hence this quantity provides a non-trivial
scaling test. Some technical details are collected in appendices A--D.

\section{A new parametrization of the FP action for
  SU(3) lattice gauge theory}\label{chap:newParametrization}

\subsection{Introduction}

In this section we present a new ansatz for the parametrization which is very
general and flexible, and which allows to parametrize the FP action using more
and more couplings without any further complications. The approach we use is
building plaquettes from the original gauge links as well as from smeared (``fat'')
links.  In this manner we are able to reproduce the classical properties of
the FP action better than with the loop parametrization.

The new ansatz is motivated by the success of using fat links in simulations
with fermionic Dirac operators
\cite{Blum:1997uf,DeGrand:1998pr,DeGrand:1999gp,Bernard:1999kc,Stephenson:1999ns,Bernard:1999xx}.
Fat links are gauge links which are locally smeared over the lattice. In this
way the unphysical short-range fluctuations inherent in the gauge field
configurations are averaged out and lattice artifacts are reduced
dramatically.

As mentioned above, earlier parametrizations of FP actions were based on
powers of the traces of loop products along generic closed paths. Restricting
the set of paths to loops up to length 8 which are fitting in a $2^4$
hypercube, one is still left with 28 topologically different loops
\cite{DeGrand:1995jk}, some of them having a multiplicity as large as 384.  In
earlier production runs only loops up to length 6 (and their powers) have been
used because including length 8 loops increases the computational cost by a
factor of $\sim 220$ \cite{DeGrand:1997gu}.  Note that in the loop
parametrization one needs length 8 loops to describe well small instanton
solutions \cite{DeGrand:1997gu,Farchioni_Papa:98}.

The new parametrization presented here provides a way around these problems,
although the computational overhead is still considerable. We have calculated
the expense of the parametrized FP action and compared it to the expense of an
optimized Wilson gauge code.  The computational overhead amounts to a factor
of $\sim 60$ per link update and comes mainly from recalculating the staples
in the smeared links affected by the modified link.

\subsection{The FP action}
We consider SU($N$) pure gauge theory\footnote{The following equations are
  given for general $N$, although the numerical analysis and simulations are
  done for SU(3).} in four dimensional Euclidean space-time on a periodic
lattice. The partition function is defined through
\begin{equation}
  \label{eq:partition_function}
  Z(\beta) = \int dU e^{-\beta {\cal A}(U)} ,
\end{equation}
where $dU$ is the invariant group measure and $\beta {\cal A}(U)$ is some
lattice regularization of the continuum action. We can perform a real space
RG transformation,
\begin{equation}
  \label{eq:rgt}
   e^{-\beta' {\cal A}'(V)} = \int dU \exp\left\{-\beta ({\cal A}(U) + T(U,V))\right\},
\end{equation}
where $V$ is the blocked link variable and $T(U,V)$ is the blocking kernel
defining the transformation,
\begin{equation}
  \label{eq:blocking_kernel}
  T(U,V) = - \frac{\kappa}{N} \sum_{n_B,\mu} \left(\text{Re} \text{Tr}(V_\mu(n_B)
  Q^\dagger_\mu(n_B)) - {\cal N}_\mu^\beta \right) .
\end{equation}
Here, $Q_\mu(n_B)$ is a $N \times N$ matrix representing some mean of 
products of link variables $U_\mu(n)$ connecting the sites $2 n_B$
and $2 (n_B +\hat\mu)$ on the fine lattice and ${\cal N}_\mu^\beta$ 
is a normalization constant ensuring the invariance of the partition 
function.
By optimizing the parameter $\kappa$, it is possible to obtain 
 an action on the coarse lattice which has a short interaction range. 
A simple choice for $Q_\mu$ is the Swendsen blocking, which contains
averaging over the 6 (long) staples along the direction $\mu$.
In ref.~\cite{Blatter:1996ti} the averaging was improved by
including more paths in $Q_\mu$.
The main idea of this block transformation is that, instead of using 
just simple staples, one additionally builds ``diagonal staples'' along 
the planar and spatial diagonal directions orthogonal to the link
direction. In this way one achieves that each link on the fine lattice 
contributes to the averaging function and the block transformation
represents a better averaging. 
In this paper we employ the RG transformation of ref.~\cite{Blatter:1996ti},
using, however, a completely new parametrization.

On the critical surface at $\beta \rightarrow \infty$, equation (\ref{eq:rgt})
reduces to a saddle point problem representing an implicit equation for the FP
action ${\cal A}^{\text{FP}}$,
\begin{equation}
  \label{eq:FP_equation}
  {\cal A}^{\text{FP}}(V) = \min_{\{U\}} \left\{{\cal A}^{\text{FP}}(U) + T(U,V)\right\}.
\end{equation}

The FP equation (\ref{eq:FP_equation}) can be studied analytically up to
quadratic order in the vector potentials \cite{Blatter:1996ti}. However, for
solving the FP equation on coarse configurations with large fluctuations one
has to resort to numerical methods, and a sufficiently rich parametrization
for the description of the solution is required.

\subsection{The parametrization}\label{sec:parametrization}

In order to build a plaquette in the $\mu\nu$-plane from smeared links we
introduce asymmetrically smeared links $W_\mu^{(\nu)}$, where $\mu$ denotes
the direction of the link and $\nu$ specifies the plaquette-plane to which
they are contributing.  This asymmetric smearing suppresses staples which lie
in the $\mu\nu$-plane relative to those in the orthogonal planes $\mu\lambda$,
$\lambda\ne\nu$.  Obviously, these two types of staples play a different role,
and we know from the quadratic approximation and the numerical evaluation of the
FP action that the interaction is concentrated strongly on the hypercube
\cite{Blatter:1996ti}.  Additional details on the smearing and the
parametrization are given in appendix \ref{app:parametrization}.  

From the asymmetrically smeared links we construct a 
``smeared plaquette variable''
\begin{equation}
  \label{wpl}
w_{\mu\nu}= \text{Re\,Tr} \left( 1-W_{\mu\nu}^{\text{pl}} \right) ,
\end{equation}
together with the ordinary Wilson plaquette variable
\begin{equation}
  \label{upl}
u_{\mu\nu}=\text{Re\,Tr}\left( 1-U_{\mu\nu}^{\text{pl}} \right) ,
\end{equation}
where
\begin{align}
  W_{\mu\nu}^{\text{pl}}(n) &= W_\mu^{(\nu)}(n) W_\nu^{(\mu)}(n+\hat \mu)
  W_\mu^{(\nu)\dagger}(n+\hat \nu) W_\nu^{(\mu) \dagger}(n) , \\
\intertext{and}
  U_{\mu\nu}^{\text{pl}}(n) &= U_\mu(n) U_\nu(n+\hat \mu)
  U_\mu^\dagger(n+\hat \nu) U_\nu^\dagger(n) . 
\end{align}

Finally, the parametrized action has the form
\begin{equation}
{\cal A}[U]= \frac{1}{N} \sum_{\mu<\nu} f(u_{\mu\nu},w_{\mu\nu}) \, ,
\end{equation}
where we choose a polynomial in both plaquette variables,
\begin{eqnarray}
f(u,w) &=& \sum_{kl} p_{kl}u^k w^l \nonumber\\
       &=& p_{10}u + p_{01}w + p_{20}u^2 + p_{11}uw + p_{02}w^2 + \ldots \,.
\end{eqnarray}
The coefficients $p_{kl}$, together with the parameters appearing in the
smearing (cf.~appendix \ref{app:parametrization}) should be chosen such that
the resulting approximation to ${\cal A}^{\text{FP}}$ is sufficiently
accurate. Note that the ansatz involves two types of parameters: the
coefficients in the asymmetric smearing enter non-linearly into the action
while the coefficients $p_{kl}$ enter linearly.

For simulations with the FP action in physically interesting regions
it is important to have a parametrization which is valid for gauge
fields on coarse lattices, i.e.~on typical rough configurations. 
We turn to this problem in the next section.

\subsection{The FP action on rough 
configurations}\label{sec:FP_action_on_rough_configurations}

The parametrization of the FP action on strongly fluctuating fields is a
difficult and delicate problem. In this section we describe briefly the
procedure of obtaining a parametrization which uses only a compact set of
parameters, but which describes the FP action still sufficiently well for the
use in actual simulations. We also provide some details about the fitting
procedure employed.

In eq.~(\ref{eq:FP_equation}) the minimizing field $U=\overline{U}(V)$ on the
fine lattice is much smoother than the original field $V$ on the coarse
lattice -- its action density is $30-40$ times smaller.  This allows an
iterative solution of eq.~(\ref{eq:FP_equation}) as follows.  First one
chooses a set of configurations $V$ with sufficiently small fluctuations such
that for the corresponding fine field $\overline{U}(V)$ one can use on the
rhs.~an appropriate starting action (the Wilson action, or better the action
${\cal A}_0$, eqs.~(\ref{rgt3}) and (\ref{plaq}) which describes well the FP
action in the quadratic approximation).  The values ${\cal A}^{\text{FP}}(V)$
obtained this way have to be approximated sufficiently accurately by choosing
the free parameters in the given ansatz.  Once this is done, one considers a
new, rougher set of configurations $V$ for which the fields $\overline{U}(V)$
have fluctuations within the validity range of the previous parametrization
and repeats the procedure described.  After 3 such steps one reaches
fluctuations typical for a lattice spacing $a\approx 0.2 - 0.3\text{ fm}$.

In previous works only the action values ${\cal A}^{\text{FP}}(V)$ have been
fitted in order to optimize the parameters.  Here we extend the set of
requirements by including into the $\chi^2$-function to be minimized the
derivatives of the FP action with respect to the gauge links in a given colour
direction $a=1,\ldots,N^2-1$,
\begin{equation}
  \label{eq:gauge_link_derivative}
  \frac{\delta {\cal A}^{\text{FP}}(V)}{\delta V_\mu^a(n)}\,.
\end{equation}
Note that these derivatives are easily calculated from the minimizing
configuration since they are given simply by $\delta T(U,V)/ \delta
V_\mu^a(n)$ at $U=\overline{U}(V)$.  In this way one has $4(N^2-1)V$ local
conditions for each configuration $V$ instead of a single global condition,
the total value ${\cal A}^{\text{FP}}(V)$.  (In addition, a good description
of local changes is perhaps more relevant in a Monte Carlo simulation with
local updates.)  An important test for the flexibility of the parametrization
is whether both the requirements for fitting the derivatives and the action
values can be met at the same time. This is indeed the case.

\begin{figure}[hbt]
\begin{center}
\includegraphics[angle=-90,width=9cm]{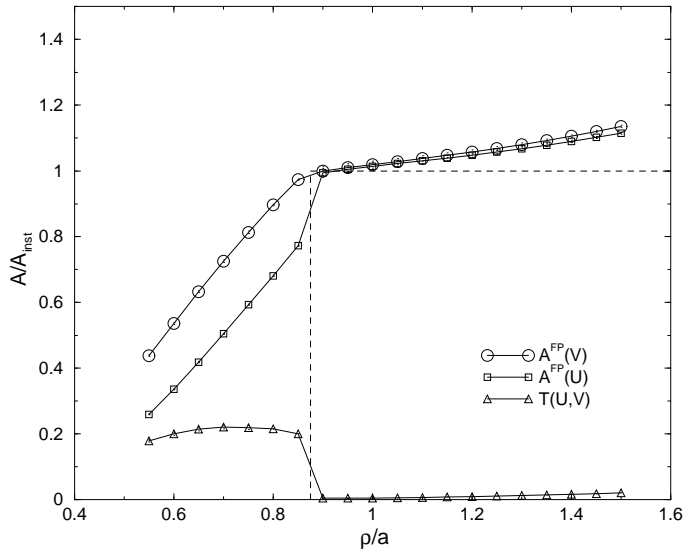}
\end{center}
\caption{{}Action values for SU(2) single instanton solutions 
  $V$ on a $6^4$ lattice in units of the continuum instanton action
  $A_{\text{inst}}$ as a function of the instanton radius $\rho/a$.
  Note, that $A^{\text{FP}}(V)=A^{\text{FP}}(U)+T(U,V)$,
  where $U$ is the minimized configuration on the fine lattice.}
\label{fig:min_hc_act}
\end{figure}

For addressing questions concerning topology it is crucial that the
pa\-ra\-met\-ri\-za\-tion describes accurately enough the exactly
scale-invariant (lattice) instanton solutions
\cite{DeGrand:1996ih,DeGrand:1996zb,DeGrand:1997gu,Farchioni_Papa:98}.  For
this purpose we generate sets of SU(2) single instanton configurations
embedded in SU(3) on a $12^4$ lattice with instanton radius $\rho/a$ ranging
from 3.0 down to 1.1.  We then block the configurations down to a $6^4$
lattice (using the blocking which defines the RG transformation).  As can be
seen from figure \ref{fig:min_hc_act}, these are solutions of the FP equations
of motion for radii $\rho/a \gtrsim 0.9$.\footnote{This is only approximately
  true: one should start from a very fine lattice and perform many
  blocking steps, and furthermore the periodic boundary conditions also violate
  (locally) the FP equations of motion.} Note that for $\rho/a \lesssim 0.9$
the quantity $T(U,V)$ becomes non-zero, indicating that instantons of that
size ``fall through the lattice'', i.e.~they are no longer solutions. The
deviation from scaling at larger radii seen in figure \ref{fig:min_hc_act} is
due to the discontinuity at the boundary of the periodic lattice and is under
control.

In the final step, we first fit the derivatives on $\sim 50$ thermal
configurations corresponding roughly to a Wilson critical coupling at 
$N_\tau \approx 2$, $\beta_c^W \approx 5.1$. In the following the
non-linear parameters (defining the asymmetric smearing) are kept fixed, while
we include in addition the action values and the derivatives of $\sim 75$
thermal configurations at $\beta^{\text{FP}}=$ 2.8, 4.0, 7.0 and the action
values of the instanton configurations.  The corresponding $\chi^2$ is then
minimized only in the linear parameters $p_{kl}$.

To assure stability of the fit we check that the $\chi^2$ is stable 
on independent configurations which are not included in the fit. 
Using high order polynomials of the plaquette variables $u$ and $w$ 
there is a danger of generating fake valleys in the
$uw$-plane (for $u$, $w$ values which are not probed by the typical
configurations and hence not restricted in $\chi^2$).
The presence of such regions is dangerous since it can force the
system to an atypical -- e.g.~antiferromagnetic -- configuration.
We find that this can be circumvented by choosing an
appropriate set of parameters. (Note that, as usually when parametrizing
the FP action, there are flat directions in the parameter space
along which the value of $\chi^2$ changes only slightly, i.e.
there is a large freedom in choosing the actual parameters.)

The smallest acceptable set of parameters consists of four non-linear
parameters describing the asymmetrically smeared links $W_\mu^{(\nu)}$ and
fourteen linear parameters $p_{kl}$ with $0 < k+l \leq 4$. The values of these
parameters are given in appendix \ref{app:param_FP} and fulfill the
correct normalization. They form the final approximation of the FP action.

Compared to the loop parametrization of ref.~\cite{Blatter:1996ti} the present
parametrization gives a deviation from the true FP action values smaller by a
factor of 2 for configurations which are typical for the range of lattice
spacing $0.03\text{ fm} \lesssim a \lesssim 0.2\text{ fm}$.  It also describes
scale invariant lattice instantons for $\rho/a \approx 1.1$ to a precision
better than 2\%. Note that this parametrization is not intended to be used on
extremely smooth configurations.  In order to be able to describe the typical
(large) fluctuations by a relatively simple ansatz we did not implement the
${\cal O}(a^2)$ Symanzik conditions (cf.~appendix \ref{app:Symanzik}) in the
last step.  In the intermediate steps (i.e.~for smaller fluctuations),
however, our parametrization (containing non-constant smearing coefficients
$\eta(x)$, $c_i(x)$) is optimized under the constraint to satisfy the ${\cal
  O}(a^2)$ Symanzik conditions as well.

In our Monte Carlo simulations we use only the final
parametrization with constant $\eta$ and $c_i$.
The parameters of the intermediate approximations to the FP action 
are not given here.

To investigate the lattice artifacts with our parametrization, we perform a
number of scaling tests, which are described in the subsequent sections.

\pagebreak

\section{The critical temperature of the deconfining phase transition}
\label{chap:deconfining_phase_transition}

\subsection{Details of the simulation}\label{critTemp:sim_details}

Using the parametrized FP action we perform a large number of simulations on
lattices with temporal extension $N_\tau=2$, $3$ and 4 at three to six
different $\beta$-values near the estimated critical $\beta_c$. Various
spatial extensions $N_\sigma /N_\tau = 2.5 \ldots 5$ are explored with the
intention of examining the finite size scaling of the critical couplings.
Configurations are generated by alternating Metropolis and overrelaxation
updates.

In the equilibrated system we measure the Polyakov loops averaged 
over the whole lattice,
\begin{equation}
  \label{eq:spatial_avg_L}
  L \equiv \frac{1}{N_\sigma^3} \sum_{\vec x} \text{Tr} 
  \prod_{t=0}^{N_\tau-1} U_4(\vec x,t),
\end{equation}
 as well as the action value of the configuration after each sweep.  
Both values are stored for later use in a spectral density reweighting 
procedure.

The details of the simulation and the run parameters are collected 
in tables \ref{tab:sim_details_N_t=2}, \ref{tab:sim_details_N_t=3}
and \ref{tab:sim_details_N_t=4}, where we list the lattice size 
together with the $\beta$-values and the number of sweeps. 
However, near a phase transition the number of sweeps is an inadequate 
measure of the collected statistics, because the resulting error is 
strongly influenced by the persistence time and the critical 
slowing down. The persistence time of one phase is defined as 
the number of sweeps divided by the observed number of flip-flops 
between the two phases \cite{Fukugita:1990yw}. 
This quantity makes sense only for $\beta$-values near the critical 
coupling $\beta_c$ and has to be taken with care: 
for the small volumes which we explore, the fluctuations within 
one phase can be as large as the separation between the two phases, 
and the transition time from one state to the other is sometimes 
as large as the persistence time itself. 
The estimated persistence time $\tau_p$ and the integrated
autocorrelation time $\tau_{\text{int}}$ of the Polyakov loop operator
are listed in the last two columns in tables \ref{tab:sim_details_N_t=2},
\ref{tab:sim_details_N_t=3} and \ref{tab:sim_details_N_t=4}. 

\subsection{Details of the analysis}\label{critTemp:analysis_details}

For the determination of the critical couplings in the thermodynamic limit we
resort to a two step procedure.  First we determine the susceptibility of the
Polyakov loops,
\begin{equation}
  \label{eq:def_Pol_loop_susc}
  \chi_L \equiv V_\sigma 
  \left( \langle|L|^2\rangle - \langle|L|\rangle^2 \right),
  \quad V_\sigma = N_\sigma^3,
\end{equation}
as a function of $\beta$ for a given lattice size and locate the position of
its maximum.  In the thermodynamic limit the susceptibility develops a delta
function singularity at a first order phase transition.  On a finite lattice
the singularity is rounded off and the quantity reaches a peak value
$\chi_L^{\text{peak}}$ at some $\beta_c(V_\sigma)$.

The critical coupling, i.e.~the location of the susceptibility peak,
is determined by using the spectral density reweighting
method, which enables the calculation of observables away from the values of
$\beta$ at which the actual simulations are performed. This method has been
first proposed in \cite{McDonald:1967,Falcioni:1982cz} and has been developed
further by Ferrenberg and Swendsen \cite{Ferrenberg:1988yz,Ferrenberg:1989ui}.

\begin{figure}[htbp]
  \begin{center}
    \includegraphics[width=6cm]{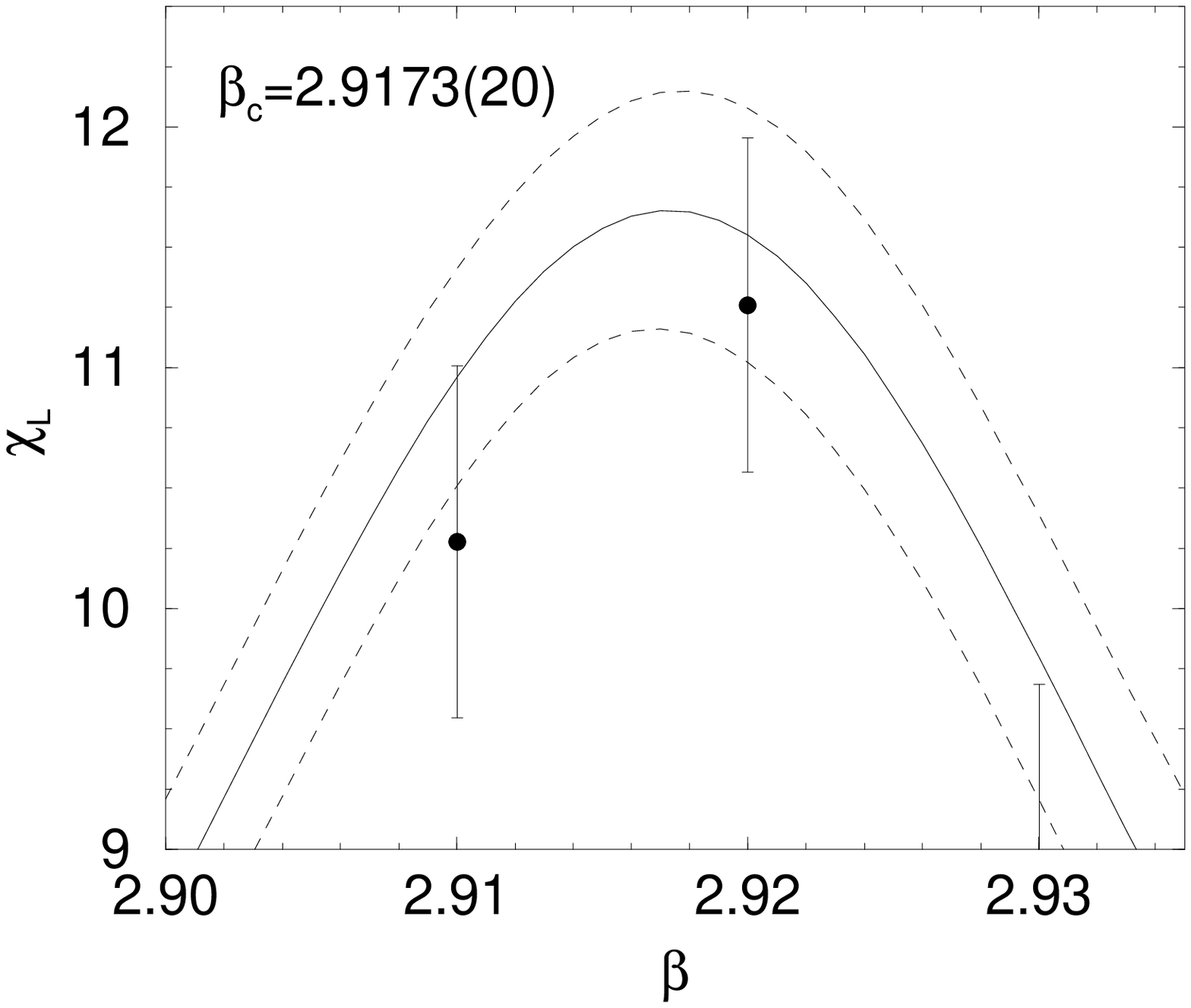}
    \includegraphics[width=6cm]{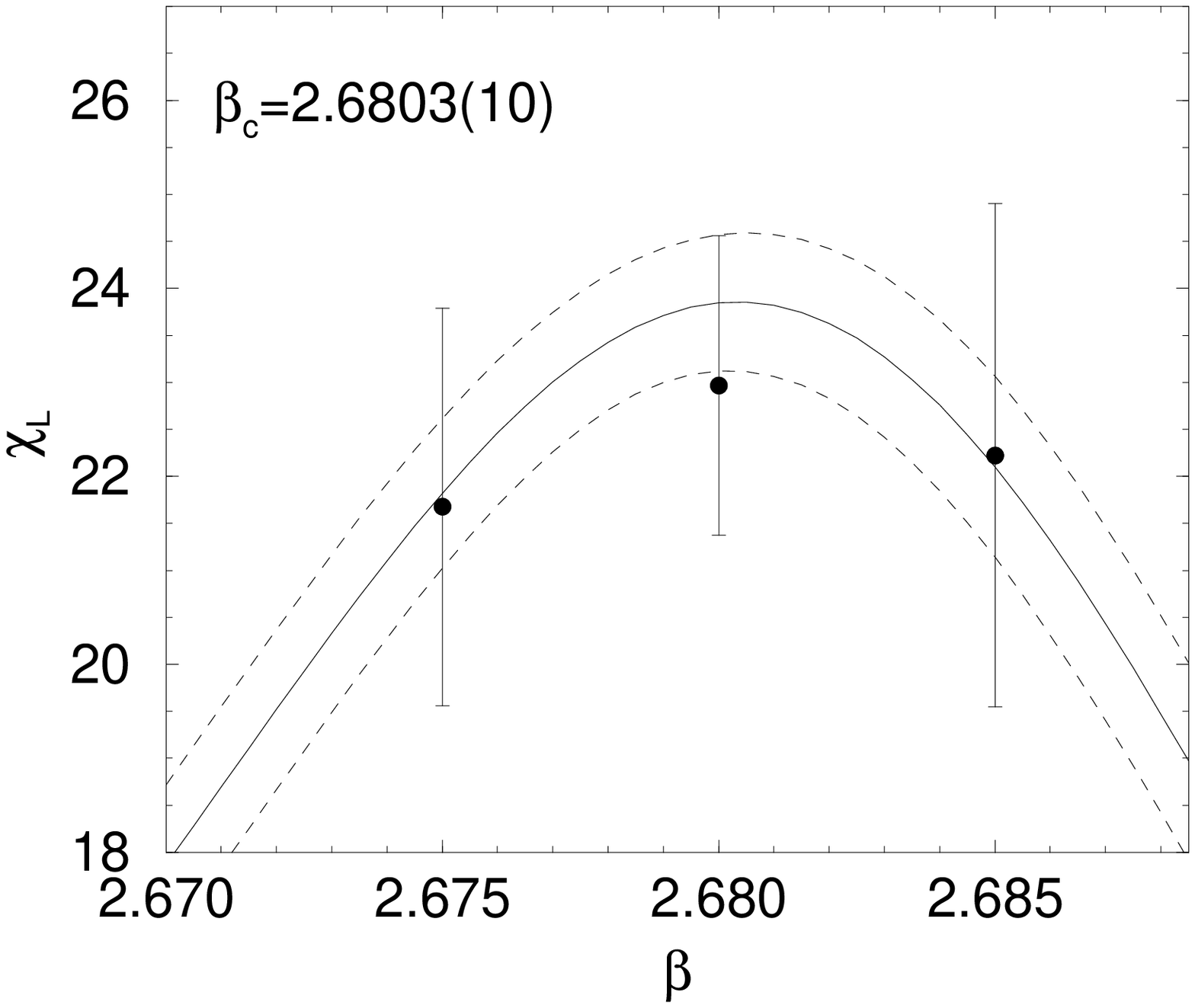}\\
    \vspace{0.3cm}
    \includegraphics[width=6cm]{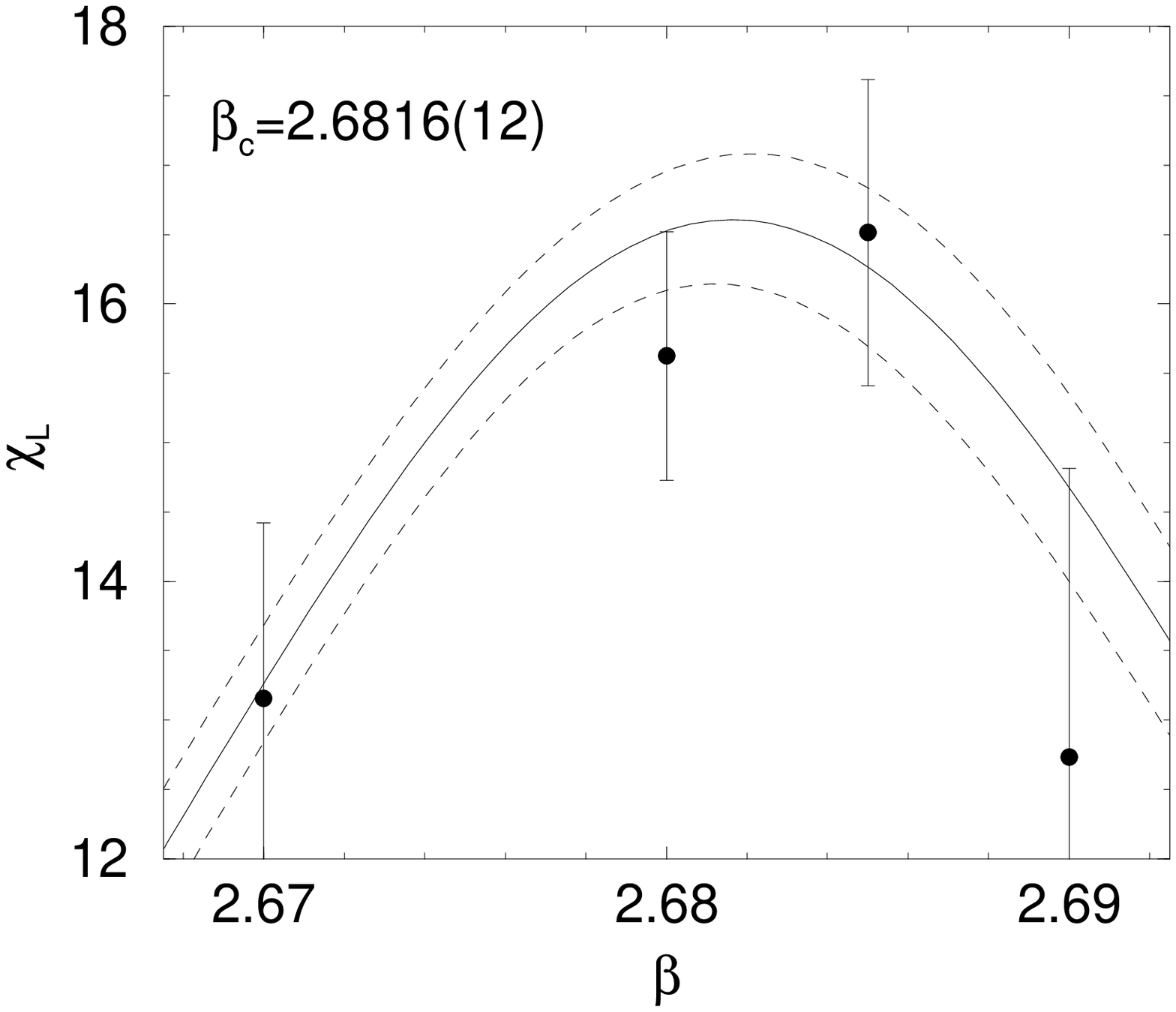}
    \includegraphics[width=6cm]{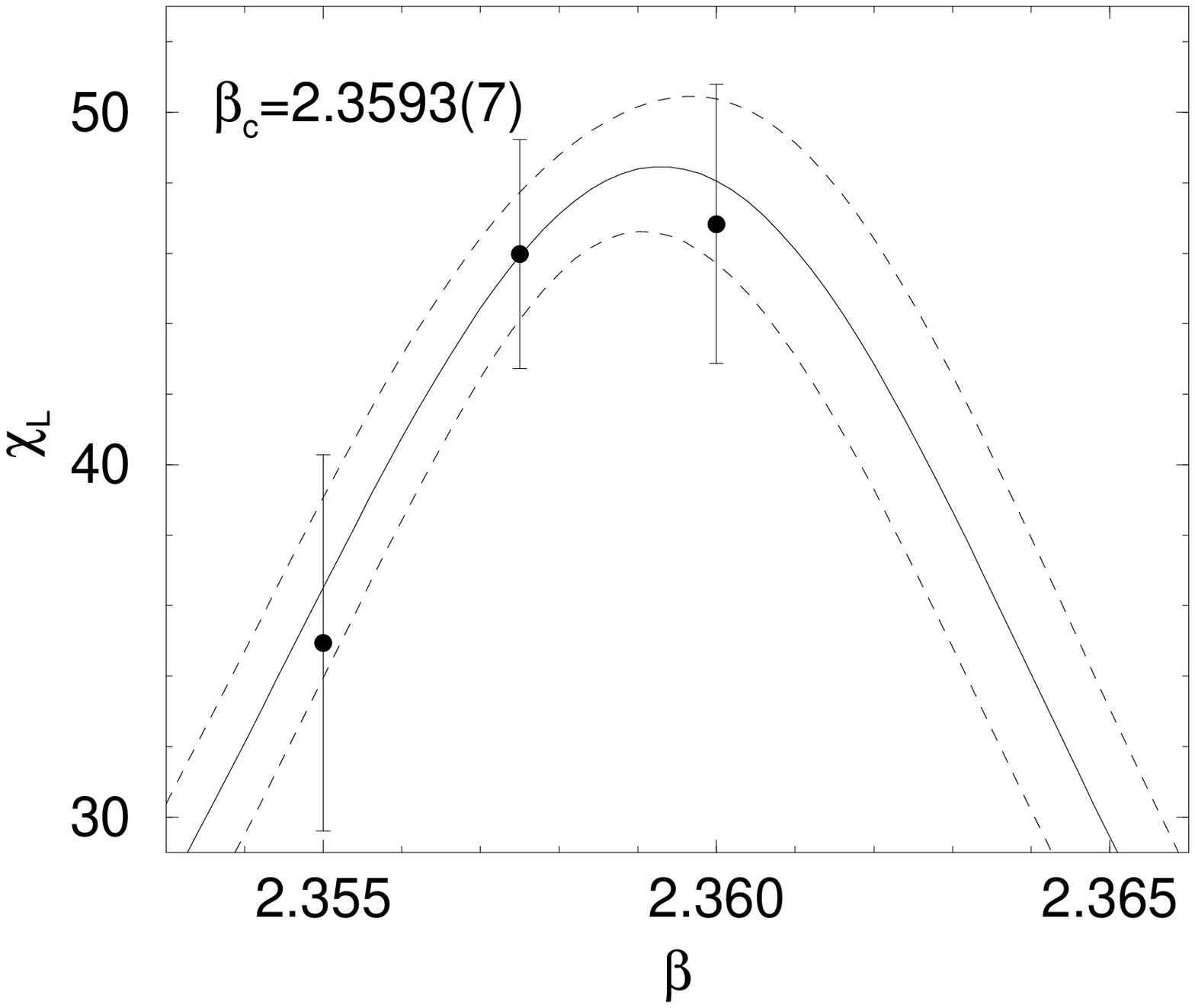}
  \caption{{}The Polyakov loop susceptibility on lattices of size 
    $4 \times 12^3$, $3 \times 12^3$, $3 \times 10^3$ and $2 \times
    10^3$. The solid curves are the interpolations using the spectral 
    density reweighting, the dashed lines show the bootstrap error bands.}
  \label{fig:Polyakov_loop_susc_peaks}
  \end{center} 
\end{figure}

In a second step we extrapolate the critical couplings for each value 
of $N_\tau$ to infinite spatial volume using the finite size scaling 
law for a first order phase transition,
\begin{equation}
  \label{eq:finite_size_scaling_beta_c}
  \beta_c(N_\tau,N_\sigma) = \beta_c(N_\tau,\infty) 
  - h \left( \frac{N_\tau}{N_\sigma}\right)^3,
\end{equation}
where $h \approx 0.1$ is considered to be a universal quantity independent of
$N_\tau$ \cite{Beinlich:1997ia}.  In figure \ref{fig:Polyakov_loop_susc_peaks}
we show the susceptibility $\chi_L$ as a function of $\beta$. The solid lines
are the interpolations obtained by the reweighting method and the dashed lines
represent the bootstrap error band estimations.  For the interpolations at a
given lattice size we use the data at all beta values listed in tables
\ref{tab:sim_details_N_t=2}--\ref{tab:sim_details_N_t=4}, although the runs at
$\beta$-values far away from the critical coupling do not influence the final
result.

In table \ref{tab:finite_size_critical_couplings} we display the values of
$\beta_c(V_\sigma)$ together with their extrapolations to $V_\sigma\to\infty$
according to formula (\ref{eq:finite_size_scaling_beta_c}).

\begin{table}[htbp]
  \begin{center}
    \begin{tabular}[c]{|cccc|}
      \hline 
        $N_\sigma$ & $\beta_c(N_\tau=2)$ & $\beta_c(N_\tau=3)$ &
        $\beta_c(N_\tau=4)$ \\
      \hline
        6        & 2.3552(24) &            &            \\
        8        & 2.3585(12) & 2.6826(23) &            \\
        10       & 2.3593(7)  & 2.6816(12) & 2.9119(31) \\
        12       &            & 2.6803(10) & 2.9173(20) \\
        14       &            &            & 2.9222(20) \\
      \hline
        $\infty$ & 2.3606(13) & 2.6796(18) & 2.9273(35) \\
         $h$     &  0.14(9)   & -0.05(7)   & 0.25(9)   \\
      \hline   
    \end{tabular}
    \caption{Results of the critical couplings $\beta_c$ from 
      the peak location of the Polyakov loop susceptibility and 
      the corresponding infinite volume limit obtained according 
      to relation (\ref{eq:finite_size_scaling_beta_c}). 
      The finite size scaling constant $h$ is also given.}
    \label{tab:finite_size_critical_couplings}
  \end{center}
\end{table}

\clearpage

\section{Scaling of the static quark-antiquark 
potential}\label{chap:scaling_properties}

\subsection{Introduction}
An important part of lattice simulations is the determination of 
the actual lattice spacing $a$ in order to convert dimensionless
quantities measured on the lattice into physical units. 
From the static $q\bar{q}$ potential one usually determines $a$
in units of the string tension $\sigma$ or in units of the hadronic 
scale $r_0$.
Using the string tension to set the scale is plagued by
some difficulties.
Since the noise/signal ratio increases rapidly with increasing $r$,
the part of the potential $V(r)$ from which the linear behaviour
$\sigma r$ has to be extracted is measured with larger statistical 
errors.
Further, due to the fact that the excited string has a small energy 
gap at large $q\bar{q}$ separations, it is difficult to resolve 
the ground state, and this can lead to a systematic error
which increases the obtained value of the string tension.

The use of the quantity $r_0$ circumvents these problems. 
In \cite{Sommer:1994ce} a hadronic scale $r_c$ has been introduced through
the force $F(r)$ between static quarks at intermediate
distances $0.2 \, \text{fm} \lesssim r \lesssim 1.0 \, \text{fm}$, 
where one has best information available from phenomenological
potential models \cite{Richardson:1979bt,Eichten:1980ms}
and where one gets most reliable results on the lattice.
One has
\begin{equation}
  \label{eq:def_hadronic_scale}
  r_c^2 V'(r_c) = r_c^2 F(r_c) = c, 
\end{equation}
where originally \cite{Sommer:1994ce} $c=1.65$ has been chosen
yielding a value
$r_0 \approx 0.49 \, \text{fm} = (395 \, \text{MeV})^{-1}$ 
from the potential models.
However, on coarse lattices also this alternative way of setting 
the scale has its ambiguities as will be discussed 
in section \ref{subsec:r0_Tc}. 

Referring to precision measurements of the low-energy
reference scale in quenched lattice QCD with the Wilson action
\cite{Bali:1992ab,Edwards:1997xf,Guagnelli:1998ud} we collect values
for $c$ and $r_c$ in table \ref{tab:hadronic_scale}. 
The first line is calculated from data in \cite{Bali:1992ab}
while the two last lines are taken from \cite{Edwards:1997xf}.
\begin{table}[htbp]
  \begin{center}
    \begin{tabular}{|cc|}
      \hline
      $r_c/r_0$ & c \\
      \hline
       0.662(1)     & 0.89 \\
       1.00         & 1.65 \\
       1.65(1)      & 4.00 \\
       2.04(2)      & 6.00 \\
      \hline
    \end{tabular}
    \caption{{}Parameter values for the determination of the hadronic
      scale through eq.~(\ref{eq:def_hadronic_scale}).}
    \label{tab:hadronic_scale}
  \end{center}
\end{table}

The scaling of our parametrized FP action is examined by measuring the
static $q\bar{q}$ potential and comparing the
quantity $r_0\left( V(r)-V(r_0)\right)$ versus $r/r_0$ at several values of
$\beta$. 

From the potential one can calculate $r_0$ and the effective string
tension $\sigma$. Finally one can test the scaling of the dimensionless
combinations $r_0 T_c$, $T_c/\sqrt{\sigma}$ and $r_0 \sqrt{\sigma}$.
This will be described in the next sections.

The scaling checks will be pushed to the extreme by exploring the behaviour of
the FP action on coarse configurations with very large fluctuations
corresponding to $a \approx 0.3$ fm.  This situation is not relevant for
practical applications, and it becomes indeed more and more difficult to
measure physical quantities due to the very small correlation length and
rapidly vanishing signals.  Nevertheless, it is still interesting to
investigate this situation in order to estimate the region in which the classical
approximation to the renormalization group trajectory is still valid.

For the standard plaquette action the static potential on coarse lattices
shows strong violations of rotational symmetry
\cite{DeGrand:1995jk,Lang:1982tj,Alford:1995hw} and, before fitting it with a
function of $r$, usually an empirical term (the lattice Coulomb potential
minus $1/r$) is subtracted with an appropriate coefficient.  On the contrary,
for the FP action of ref.~\cite{Blatter:1996ti}, due to the proper choice of
the RG transformation the resulting potential is (practically) rotational
invariant\footnote{Note that to cure the rotational invariance of the
  potential one has to improve not only the action but also the operators. For
  a more ``rotational invariant'' blocking the potential shows less violations
  of rotational symmetry.}.  Here we do not aim at testing the rotational
invariance of the potential but rather at determining $r_0$ and $\sigma$.

\subsection{Details of the simulation}\label{sec:potential_simulation_details}
We perform simulations with the FP action at six different $\beta$-values, of
which three correspond to the critical couplings determined in section
\ref{chap:deconfining_phase_transition}.  Configurations are updated by
alternating Metropolis with overrelaxation sweeps.  The spatial
extent of the lattices is chosen to be at least $\sim 1.5$ fm, based on
observations in \cite{Bali:1992ab,Edwards:1997xf,Bali:1993ru}. 
Table \ref{run_parameters_for_potentials} contains the values of the
couplings, the lattice sizes and the number of measurements.

In order to enhance the overlap with the physical ground state of the
potential we exploit smearing techniques.
The smoothing of the spatial links has the
effect of reducing excited-state contaminations in the correlation
functions of the strings in the potential measurements.
The operators which we measure in the simulations are constructed
using the spatial smearing of \cite{Albanese:1987ds}. It consists
of replacing every spatial link
$U_j(n)$, $j=$ 1, 2, 3 by itself plus a sum of its neighbouring spatial
staples and then projecting back to the nearest element in the SU(3)
group:
\begin{eqnarray}
  {\cal S}_1 U_j(x) & \equiv & {\cal P}_{\text{SU(3)}} 
   \Big\{ U_j(x) + \lambda_s \sum_{k \neq j}
  (U_k(x) U_j(x+\hat k) U_k^\dagger(x+\hat j)  \\
 & &  \hspace{2cm}  + U_k^\dagger(x-\hat k)
  U_j(x-\hat k) U_k(x-\hat k+\hat j))  \Big\}. \nonumber 
\end{eqnarray}
Here, ${\cal P}_{\text{SU(3)}} Q$ denotes the unique projection onto the SU(3)
group element $W$ which maximizes $\text{Re} \text{Tr}(W Q^\dagger)$ for an
arbitrary $3 \times 3$ matrix $Q$. The smeared and SU(3) projected link ${\cal
  S}_1 U_j(x)$ retains all the symmetry properties of the original link $
U_j(x)$ under gauge transformations, charge conjugation, reflections and
permutations of the coordinate axes.  The set of spatially smeared links
$\{{\cal S}_1 U_j(x) \}$ forms the spatially smeared gauge field
configuration.  An operator $\cal O$ which is measured on a $n$-times
iteratively smeared gauge field configuration is called an operator on
smearing level ${\cal S}_n$, or simply ${\cal S}_n {\cal O}$.  In the
simulation of the static $q\bar{q}$ potential we use smearing levels with
$n=$ 0, 1, 2, 3, 4.  The smearing parameter is chosen to be $\lambda_s=0.2$ in all
cases.

The correlation matrix of spatially smeared strings is constructed in
the following way. At fixed $\tau$ we first form smeared string operators
along the three spatial axes, connecting $\vec x$ with $\vec x + r \hat i$,
\begin{multline}
  \label{eq:smeared_spatial_link}
  {\cal S}_n V_i(\vec x, \vec x + r \hat i;\tau) = \\
  {\cal S}_n U_i(\vec x,\tau) {\cal S}_n U_i(\vec x+\hat i,\tau) \ldots 
   {\cal S}_n U_i(\vec x + (r-1) \hat i,\tau), \quad i=1,2,3,
\end{multline}
and unsmeared temporal links at fixed $\vec x$, connecting $\tau$ with
$\tau+t$,
\begin{equation}
  \label{eq:unsmeared_temporal_link}
  V_4(\tau,\tau+t;\vec x) = U_4(\vec x,\tau) U_4(\vec x,\tau+1) \ldots U_4(\vec
  x,\tau+(t-1)).
\end{equation}
Finally, the correlation matrix is given by
\begin{multline}
  \label{eq:Wilson_loop_correlation}
C_{lm}(r,t) =
\Big\langle
 \sum_{\vec x, \tau} \sum_{i=1}^3 \text{Tr} \, 
  {\cal S}_l V_i(\vec x, \vec x + r \hat i;\tau) 
  V_4(\tau,\tau+t;\vec x + r \hat i) \\
   {\cal S}_m V_i^\dagger(\vec x, \vec x + r \hat i;\tau+t) 
  V_4^\dagger(\tau,\tau+t;\vec x) \Big\rangle \,,
\end{multline}
where $\langle . \rangle$ denotes the Monte Carlo average.  
In the following the correlation matrices are analyzed as described 
in section \ref{sec:potential_analysis}.

\subsection{Details of the analysis and results}\label{sec:potential_analysis}
In order to extract the physical scale through equation 
(\ref{eq:def_hadronic_scale}) we need an interpolation of the potential 
and correspondingly the force between the quarks for arbitrary distances 
$r$. 
This interpolation of $V(r)$ is achieved by fitting an ansatz of the form
\begin{equation}
  \label{eq:potential_ansatz}
  V(r) = V_0 - \frac{\alpha}{r} + \sigma r
\end{equation}
to the measured potential values. 
  
We determine the scale in two steps. First we employ the variational
techniques described in appendix \ref{sec:var_tech} using the correlation
matrix defined in eq.~(\ref{eq:Wilson_loop_correlation}) for a given
separation $r$.  This method results in a linear combination of string
operators ${\cal S}_n V$, $n =0,\ldots,4$, which projects sufficiently well
onto the ground state of the string, i.e.~eliminates the closest excited
string states.  We then build a $\chi^2$ function using the covariance matrix
which incorporates correlations between $C_{lm}(t)$ and $C_{l'm'}(t')$.  Based
on effective masses and on the $\chi^2$ values we choose a region
$t_{\text{min}}(r) \le t \le t_{\text{max}}(r)$.  (Too small $t$ values can
distort the results due to higher states which are not projected out
sufficiently well, while too large $t$ values are useless due to large
errors.)  In this window we fit the ground state correlator by the exponential
form $Z(r) \exp(-t V(r))$, checking the stability under the variation of
different parameters of the procedure.  The results of these fits are
collected in table \ref{tab:pot_values_all_betas}.

A straightforward strategy is to fit the obtained values of $V(r)$ by the
ansatz (\ref{eq:potential_ansatz}).  
However, one can decrease the errors on $\alpha$
and $\sigma$ by exploiting the fact that the errors of $V(r)$ at different
values of $r$ are correlated. Therefore, in the second step, we use the
projectors to the ground state of the string for each $r$, obtained in the
first step, and calculate the ground state correlator $\bar{C}(r,t)$ from
the correlation matrix $C_{lm}(r,t)$.  After this we estimate the covariance
matrix $\text{Cov}(r,t;r',t')$ from the bootstrap samples of $\bar{C}(r,t)$
and use this to build a $\chi^2$ function, fitting $\bar{C}(r,t)$ with the
expression $Z(r) \exp(-t (V_0 - \alpha/r + \sigma r))$.  We use the fit range
$t_{\text{min}}(r) \le t \le t_{\text{max}}(r)$ determined in the first step.
The fit range in $r$ is determined by examining the $\chi^2$ values and 
the stability of the fitted parameters. The results of these fits are given 
in table \ref{tab:pot_fit_results}.

Having in hand a global interpolation of the static potential for each
$\beta$-value, we are able to determine the hadronic scale $r_0$ in units of
the lattice spacing through eq.~(\ref{eq:def_hadronic_scale}). The value of
$c$ is chosen appropriate to the coarseness of the lattice and the fit range
in $r$.

In addition, we repeat the second step, but restricting this time the values
of $r$ to values close to $r_c$ to have a local fit to $V(r)$
\cite{Sommer:1994ce,Guagnelli:1998ud}.  This fit is used then again to
determine $r_c/a$ (and accordingly $r_0/a$) from the relation
(\ref{eq:def_hadronic_scale}).

The final results for $r_0/a$ are listed in table \ref{tab:r0_from_local_fit}
where the first error denotes the purely statistical error.  The second one
represents an estimate of the systematic error and marks the minimal and
maximal value of $r_0/a$ obtained with different local fit ranges and
different reasonably chosen values of $c$. These ambiguities are discussed in
detail in section \ref{subsec:r0_Tc}.

\begin{table}[htbp]
  \begin{center}
    \renewcommand{\arraystretch}{1.3}
    \begin{tabular}{|ccl|}
\hline\vspace{-0.05cm}
      $\beta$ & $N_\tau$ &  \phantom{2cm}$r_0/a$     \\
\hline
        3.400 &          &  $4.833(39)(\genfrac{}{}{0pt}{}{+18}{-22})$ \\ 
        3.150 &          &  $3.717(23)(\genfrac{}{}{0pt}{}{+16}{-17})$ \\
        2.927 &    4     &  $2.969(14)(\genfrac{}{}{0pt}{}{+5}{-14})$  \\
        2.860 &          &  $2.740(10)(\genfrac{}{}{0pt}{}{+17}{-31})$ \\
        2.680 &    3     &  $2.237(7)(\genfrac{}{}{0pt}{}{+11}{-33})$  \\
        2.361 &    2     &  $1.500(5)(\genfrac{}{}{0pt}{}{+29}{-14})$  \\
\hline
      \end{tabular}
    \caption{{}The hadronic scale $r_0/a$ determined from local fits
    to the potential. The first error denotes the statistical error
    and the second is the estimate of the systematic error.}
    \label{tab:r0_from_local_fit}
  \end{center}
\end{table}

\begin{figure}[htbp]
  \begin{center}
    \includegraphics[width=9cm]{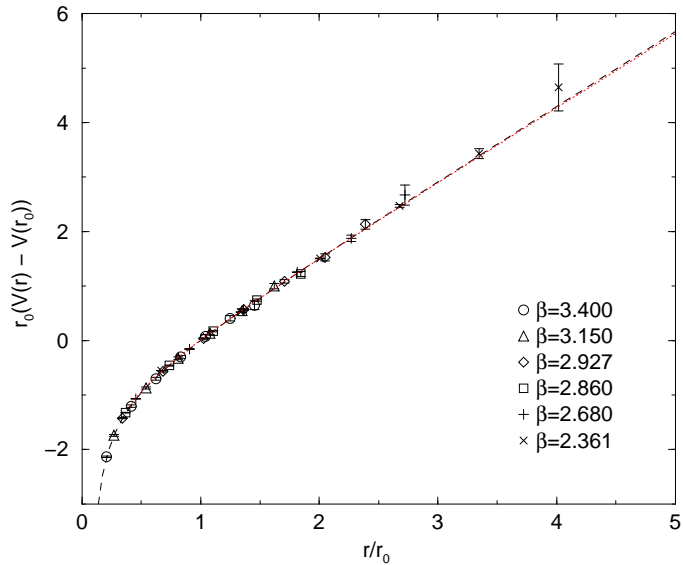}
    \caption{{}Scaling of the static $q\bar{q}$ potential $V(r)$
    expressed in terms of the hadronic scale $r_0$. The unphysical 
    constant $r_0 V(r_0)$ is subtracted for each lattice spacing such
    that the curves match at $r/r_0=1$. 
    The dashed line is a fit of the form (\ref{eq:potential_ansatz}) 
    to the data. 
    The dotted line, which practically coincides with our fit, 
    is from  \cite{Juge:1997nc}.}
    \label{fig:all_pot}
  \end{center}
\end{figure}

In figure \ref{fig:all_pot} we display the potential values.  The dashed line
is obtained by a simultaneous fit to all the data respecting the previously
chosen fit ranges in $r$.  The dotted line representing the result of
\cite{Juge:1997nc} (obtained on anisotropic lattices using an improved action)
is hardly distinguishable from our dashed line.

\begin{figure}[htbp]
  \begin{center}
    \includegraphics[width=9cm]{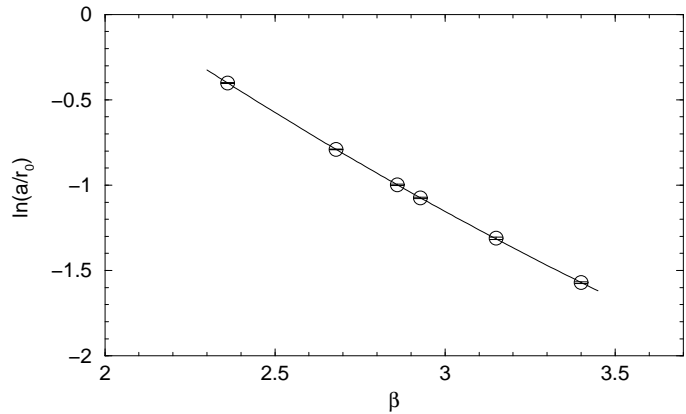}
    \caption{{}The measured data points of $\ln(a/r_0)$ (circles) and
    their phenomenological description in terms of a polynomial
    quadratic in $\beta$ (solid line). The plotted points are the
    values of $r_0/a$ from local fits.}
    \label{fig:r0vsbeta}
  \end{center}
\end{figure}

It could be useful to have an empirical interpolating formula connecting the
lattice spacing to the bare coupling.  (Analogous fits for the Wilson action
are given in refs.  \cite{Edwards:1997xf,Guagnelli:1998ud}.)  The expression
\begin{equation}
  \label{eq:interpolating_ansatz}
  \ln(a/r_0) =
    -1.1622(24) -1.0848(95) (\beta - 3) + 0.156(17)(\beta - 3)^2
\end{equation}
describes well the data points in the range $2.361 \le \beta \le 3.4$. The fit
is shown in figure \ref{fig:r0vsbeta}.

\section{{}Scaling of the critical temperature and 
  \boldmath$r_0 \sqrt{\sigma}$ }
To further study the scaling properties of the FP action we examine the 
dimensionless combinations of physical quantities $T_c/\sqrt{\sigma}$,
$r_0 T_c$  and $r_0 \sqrt{\sigma}$. 

In this section we present and discuss the results for the FP action and
compare them to results obtained for the Wilson action and different
improved actions whenever it is possible.

\subsection{{}\boldmath$T_c/\sqrt{\sigma}$ }\label{sec:sigmaTc}
Let us first look at the ratio $T_c/\sqrt{\sigma}$, the deconfining
temperature in terms of the string tension\footnote{Here and in the following
  we refer to the quantity $\sigma$ obtained from the three parameter fits of
  the form (\ref{eq:potential_ansatz}) to the static potential as the
  (effective) string tension.}.

\begin{table}[htbp]
  \begin{center}
    \begin{tabular}{|lcl|}
\hline
 action   & $\beta$  & $T_c/\sqrt{\sigma}$ \\
\hline
FP action                         & 2.927    & 0.624(7)  \\
                                  & 2.680    & 0.622(8)  \\
                                  & 2.361    & 0.628(11) \\    
\hline
Wilson   \cite{Beinlich:1997ia}   & $\infty$ & 0.630(5)  \\
$1\times2$ \cite{Beinlich:1997ia} & $\infty$ & 0.634(8)  \\
DBW2  \cite{deForcrand:1999bi}    & $\infty$ & 0.627(12) \\
Iwasaki \cite{Iwasaki:1997sn}     & $\infty$ & 0.651(12) \\
Bliss \cite{Bliss:1996wy}         & $\infty$ & 0.659(8) \\
\hline
    \end{tabular}
    \caption{{}Results of the deconfining temperature in units of the
    string tension obtained with the FP action and continuum values
    for different other actions. }\label{tab:sigmaTc}
  \end{center}
\end{table}

In table \ref{tab:sigmaTc} we collect all available continuum
extrapolations together with the results for the FP action.
The data obtained with the Wilson action is taken from
\cite{Beinlich:1997ia} where they use the $T_c$ values at $N_\tau=4$
and 6 from \cite{Boyd:1996bx} and extrapolate finite volume data for
$T_c$ at $N_\tau=8$ and 12 from \cite{Boyd:1996bx} to infinite
volume. For the value of $\sigma$ they use the string tension
parametrization given in \cite{Edwards:1997xf}. The data for the 
$1\times 2$ tree level improved action is again taken from
\cite{Beinlich:1997ia}. The data denoted by RG improved action is
obtained with the Iwasaki action \cite{Iwasaki:1985we} and is taken
from \cite{Iwasaki:1997sn}. We also include the
    value by Bliss et al.~\cite{Bliss:1996wy} from a tree level and
    tadpole improved action.  Finally we quote the results from the
QCD-TARO collaboration \cite{deForcrand:1999bi} obtained with the DBW2
action\footnote{DBW2 means "doubly blocked from Wilson in two coupling
space".}. The extrapolations to the continuum stem from \cite{Teper:1998kw}
where a careful reanalysis has been done.

For extracting the string tension we follow a simple approach.
As described above, we perform fits to the on-axis potential
values only and therefore we are limited to a small number of different
fitting ranges.
Nevertheless, the values of $\sigma$ obtained this
way and quoted in table \ref{tab:pot_fit_results} are stable and vary
only within their statistical errors over the sets of sensibly
considered fit ranges. However, the error on $\sigma$ changes
considerably, i.e.~up to a factor of 5, depending on whether distance
$r=1$ is taken into account or not.  Just to play safe we neglect
distance $r=1$ in the fits, even if the $\chi^2$ would allow it.

\begin{figure}[htbp]
  \begin{center}
    \includegraphics[width=9cm,angle=-90]{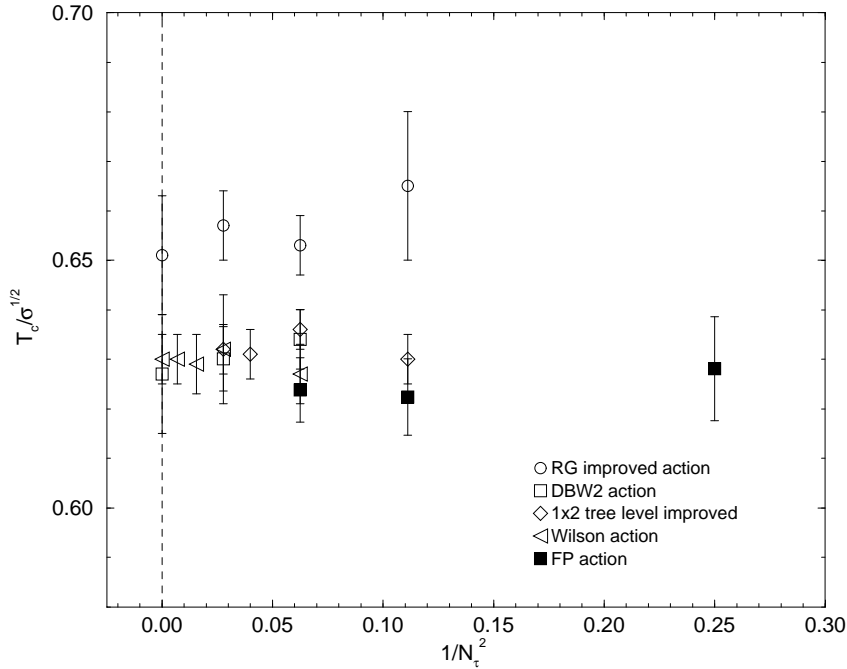}
    \caption{{}$T_c/\sqrt{\sigma}$ vs.~$1/N_\tau^2$ for different
    actions.}
    \label{fig:sigmaTc}
  \end{center}
\end{figure}

The values are displayed in figure \ref{fig:sigmaTc} together with the
data as mentioned above. Our data is compatible within one standard
deviation with the continuum extrapolation of the Wilson data and we
observe scaling of the FP action within the statistical errors over
the whole range of coarse lattices corresponding to values
of $N_\tau=$ 2, 3 and 4.

\subsection{{}\boldmath$r_0 T_c$} \label{subsec:r0_Tc}
Unfortunately, precise determinations of $r_0/a$ are missing in the literature
except for the Wilson action \cite{Edwards:1997xf,Guagnelli:1998ud} and, in
contrast to $T_c/\sqrt{\sigma}$, we are not able to compare our data to other
actions such as the Iwasaki, DBW2 or the $1 \times 2$ tree level improved
action. In fact, the determination of $r_0/a$ is a delicate issue and
systematic effects due to different methods of calculating the force can be
sizeable. Due to the fact that extracting the derivative of the potential from
a discrete set of points is not unique, the intrinsic systematic uncertainty
is not negligible at intermediate and coarse lattice spacings $a \gtrsim 0.15$
fm.  For example, in an accurate scale determination of the Wilson gauge
action in \cite{Edwards:1997xf} the authors quote a value of $r_0/a=2.990(24)$
at $\beta_W=5.7$. This is to be compared with $r_0/a=2.922(9)$ of
ref.~\cite{Guagnelli:1998ud} for the same action and $\beta$-value.  In view
of the claim in \cite{Edwards:1997xf} to have included all systematic errors
and the high relative accuracy ($\sim 0.3 \%$) of the data in
\cite{Guagnelli:1998ud}, this systematic difference seems to be a serious
discrepancy.  Even on finer lattices there are ambiguities: at $\beta_W=6.2$
the authors of \cite{Guagnelli:1998ud} obtain $r_0/a=7.38(3)$, while in
\cite{Bali:1997bj} a value of $r_0/a=7.29(4)$ is quoted.

\begin{figure}[htbp]
  \begin{center}
    \includegraphics[width=9cm,angle=-90]{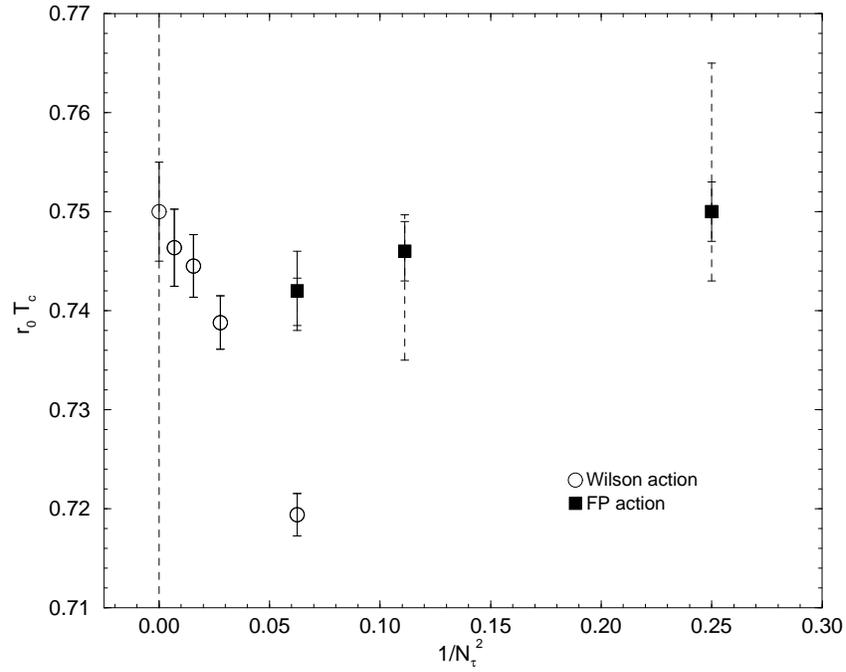}
    \caption{{}$r_0 T_c$ vs.~$1/N_\tau^2$ for the Wilson and the FP
    action. The empty circles represent data from measurements with
    the Wilson action and the filled squares denote the results
    obtained with the FP action. The solid error bars show the purely
    statistical error, while the dashed ones indicate the systematic error
    from the ambiguities in determining the force on coarse lattices.}
    \label{fig:r0Tc}
  \end{center}
\end{figure}

In that sense our results concerning $r_0 T_c$ have to be taken with
appropriate care.  In table \ref{tab:final_r0Tc} we collect the data
for $r_0 T_c$ from our measurements with the FP action together with
the data from measurements with the Wilson action.  
The critical couplings corresponding to $N_\tau=$ 4, 6, 8 and 12 are
taken from \cite{Beinlich:1997ia} while the values for $r_0/a$ are
from the interpolating formula in \cite{Guagnelli:1998ud}. The quoted
errors are purely statistical. The continuum value is our own
extrapolation obtained by performing a fit linear in the leading
correction term $1/N_\tau^2$ and discarding the data point at
$N_\tau=4$.  Finally, the values are plotted in figure \ref{fig:r0Tc}
for comparison.
\begin{table}[htbp]
  \begin{center}
    \begin{tabular}{|c|cc|}
      \hline
       $N_\tau$ & Wilson action & FP action \\
      \hline 
       2    &          &  0.750(3) \\
       3    &          &  0.746(3) \\
       4    & 0.719(2) &  0.742(4) \\
       6    & 0.739(3) &           \\
       8    & 0.745(3) &           \\
       12   & 0.746(4) &           \\
       $\infty$& 0.750(5) &         \\
      \hline
    \end{tabular}
    \caption{{}Results for the critical temperature in terms of the
    hadronic scale, $r_0 T_c$, from measurements with the Wilson
    action and the FP action.} 
    \label{tab:final_r0Tc}
  \end{center}
\end{table}

The Wilson action shows scaling violation for $r_0 T_c$ of about $4\%$
at $N_\tau=4$, while at $N_\tau=6$ it is already smaller than about
$1.5 \%$. In that sense this quantity provides a high precision
scaling test and thus a very accurate computation of the low-energy
reference scale $r_0/a$ on the $0.5 \%$ level is of crucial
importance. The lack of data for different actions is an indication
that this is indeed a difficult task. Although the required statistics
is in principle accessible to us, we do not have full control over the
systematic ambiguities in the calculation of $r_0/a$ on the required
accuracy level. Nevertheless we observe in principle excellent scaling
within $1 \%$ or two standard deviations for the FP action even on
coarse lattices corresponding to $N_\tau=3$ and 2, however, this
statement is moderated in view of the large systematic uncertainties.

\subsection{{}\boldmath$r_0 \sqrt{\sigma}$}
To obtain the dimensionless product $r_0 \sqrt{\sigma}$ we use the
values of $r_0/a$ in table \ref{tab:r0_from_local_fit} obtained from
local fits and the values of $\sigma$ as determined in section
\ref{sec:sigmaTc}, where $\sigma$ is determined from the long range
properties of the potential.

In table \ref{tab:r0sigma} we collect the resulting values of $r_0
\sqrt{\sigma}$. We can extrapolate to the continuum by performing a fit linear
in $(a/r_0)^2$ and obtain $r_0 \sqrt{\sigma} = 1.193(10)$. For comparison we
calculate the data for the Wilson action from the interpolating formula for
$r_0/a$ in \cite{Guagnelli:1998ud} and the string tension parametrization in
\cite{Edwards:1997xf}. The continuum extrapolation for the Wilson data is
taken from the analysis of Teper in \cite{Teper:1998kw}.

\begin{table}[htbp]
  \begin{center}
    \begin{tabular}{|ll|ll|}
\hline
\multicolumn{2}{|c|}{Wilson action} & \multicolumn{2}{c|}{FP action} \\
$\beta$ &  $r_0 \sqrt{\sigma}$ & $\beta$ &  $r_0 \sqrt{\sigma}$ \\
\hline
5.6925  &  1.148(12) & 2.361   &  1.194(21) \\
5.8941  &  1.170(19) & 2.680   &  1.196(15) \\
6.0624  &  1.183(13) & 2.860   &  1.190(23) \\
6.3380  &  1.185(11) & 2.927   &  1.191(12) \\
        &            & 3.150   &  1.185(16) \\
        &            & 3.400   &  1.198(12) \\
$\infty$&  1.197(11) & $\infty$&  1.193(10) \\
\hline
\end{tabular}
    \caption{{}$r_0 \sqrt{\sigma}$ for the Wilson and the FP action.}
    \label{tab:r0sigma}
  \end{center}
\end{table}

\begin{figure}[htbp]
  \begin{center}
    \includegraphics[width=9cm,angle=-90]{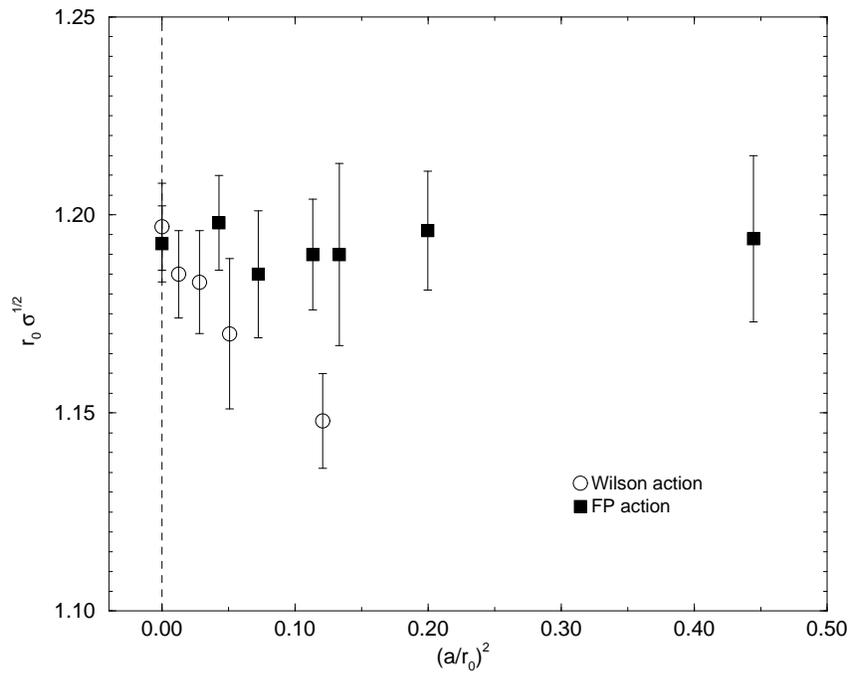}
    \caption{{}Scaling behaviour of $r_0 \sqrt{\sigma}$ for the Wilson
    action (empty circles) and the FP action (filled squares).}
    \label{fig:r0sigma}
  \end{center}
\end{figure}

Figure \ref{fig:r0sigma} shows the scaling behaviour of 
$r_0\sqrt{\sigma}$ for the Wilson action (empty circles) and the FP
action (filled squares) as a function of $(a/r_0)^2$. 
The error bars are purely statistical and are dominated by the
uncertainty from the string tension. Therefore the systematic
ambiguities present in $r_0/a$ are not visible within the error bars.

The Wilson action shows a scaling violation of about $4 \%$ at 
$\beta=5.6925$ $(N_\tau=4)$, while no scaling violation is seen for the
FP action even on lattices as coarse as $\beta=2.361$ $(N_\tau=2)$. We
would like to emphasize that this is a non-trivial result, since
$r_0/a$ and $\sqrt{\sigma} a$ are determined independently of
each other.  However, with the data presently available to us it is
difficult to extract the string tension with the accuracy needed to
see a striking difference to the Wilson action for $\beta$-values
corresponding to $N_\tau \geq 4$. This is mainly due to the lack of
measurements of the off-diagonal potential values.

\section{Glueballs}\label{chap:glueballs}

\subsection{Introduction}

Glueballs are the one-particle states of SU($N$) gauge theory.
They are characterized by the quantum numbers $J^{PC}$,
denoting the symmetry properties with respect to the O(3) rotations
(spin), spatial reflection and charge conjugation.
However, the lattice regularization does not preserve the continuous
O(3) symmetry, only its discrete cubic subgroup, therefore
the eigenstates of the transfer matrix are classified according
to irreducible representations of the cubic group.
There are five such representations:
$A_1$, $A_2$, $E$, $T_1$, $T_2$, of dimensions 1, 1, 2, 3, 3, respectively.
Their transformation properties can be described by polynomials 
in $x$, $y$, $z$ as follows: $A_1 \sim\{ 1 \}$, $A_2 \sim\{ xyz \}$, 
$E \sim\{ x^2-z^2,y^2-z^2 \}$, $T_1 \sim\{ x,y,z \}$
and $T_2 \sim\{ xy,xz,yz\}$, where $x$, $y$, $z$ are components
of an O(3) vector.
In general, an O(3) representation with spin $J$ splits into
several representations of the cubic group.
Looking at the corresponding polynomials, it is rather obvious that
the splitting starts at $J=2$: 
$(J=0) \to A_1$, $(J=1) \to T_1$, $(J=2) \to (E, T_2)$.
The full O(3) rotation symmetry is expected to be restored
in the continuum limit. This restoration manifests itself
e.g.~in the fact that a doublet $E$ and a triplet $T_2$
(for a given choice of quantum numbers $PC$) become degenerate
to form together the $J=2$ states with 5 possible polarizations.

The main obstacle in the computations of glueball masses on the
lattice is the fast decay of the signal in the correlation functions
of the gluonic excitations, due to the fact that the glueball masses
are relatively large ($m_G \gtrsim 1.6$~GeV). 
For this reason a small lattice spacing $a$ is required to follow
the signal long enough. On the other hand, the physical lattice volume
should be larger than $L\gtrsim 1.2$~fm to avoid finite size
effects. This finally results in a large $L/a$ making it hard to
obtain the statistics which is usually required. One possible way
around this dilemma is the use of anisotropic lattice actions, which
have a finer resolution in time direction, $a_\tau \ll a_\sigma$, and
where one can follow the signal over a larger number of time slices.
Although this idea is not new \cite{Ishikawa:1983xg}, it has been
revived only recently by Morningstar and Peardon
\cite{Morningstar:1997ff,Morningstar:1999rf}. Using an anisotropic
improved lattice action they investigated the glueball spectrum below
4~GeV in pure SU(3) gauge theory and improved the determinations
of the glueball masses considerably compared to previous Wilson action
calculations. Recent calculations with the Wilson action comprehend works by
the UKQCD collaboration \cite{Bali:1993fb} and the GF11 group
\cite{Chen:1994uw,Vaccarino:1999ku}.  It can be said that all three
calculations are in reasonable agreement on the masses of the two
lowest lying $0^{++}$ and $2^{++}$ glueballs.

Despite this agreement, Wilson action calculations of the $0^{++}$
glueball mass show huge lattice artifacts of around 40 \% at 
$a \approx 0.15$ fm and still 20 \% even at $a \approx 0.10$ fm. 
From this point of view the $0^{++}$ glueball mass is particularly 
interesting, besides its physical relevance, since it provides an 
excellent test object on which the scaling behaviour of different
actions can be checked and the achieved reduction of discretization
errors can be sized.  
In this sense let us emphasize that our intention here is twofold:
firstly, our calculation provides a new and independent determination
of glueball masses using FP actions, and secondly, we aim at using the
glueball spectrum, in particular the mass of the $0^{++}$ glueball, as
another scaling test of the FP action. Although we observe that the FP
action scales well in quantities like $r_0 T_c$, $T_c/\sqrt{\sigma}$ or
$r_0 \sqrt{\sigma}$, lattice artifacts could be, in principle, quite
different for other physical quantities, in particular $r_0 m_G$ or
$m_G/\sqrt{\sigma}$.

This section is organized as follows.  In subsection
\ref{sec:glueballs_sim_details} we describe the details of the
simulations including the generation of the gauge field configurations
and the measurements of the operators. The extraction of masses from
the Monte Carlo estimates of glueball correlation functions is
described in subsection \ref{sec:glueballs_analysis_details}.
Finally, subsection \ref{sec:glueball_results} contains the results of our
glueball measurements.

\subsection{Details of the simulation}\label{sec:glueballs_sim_details}

We perform simulations at three different lattice spacings in the
range $0.1 \, \text{fm} \leq a \leq 0.18 \, \text{fm}$ and volumes
between (1.4~fm)$^3$ and (1.8~fm)$^3$. The simulation parameters for our runs 
are given in table \ref{tab:glueballs_sim_details}.

The gauge field configurations are updated by performing compound
sweeps consisting of alternating over-relaxation and standard
Metropolis sweeps.

First, a rather small preliminary simulation at $\beta=2.86$ is performed.
Using the results of some pilot runs, we determine a set of five loop shapes
which have large contributions to the $A_1^{++}$ channel. Using the labelling
of Berg and Billoire \cite{Berg:1983kp} these are the length-8 loop shapes 2,
4, 7, 10, 18.  They are measured on five smearing levels ${\cal S}_n$,
$n=2,4,\ldots, 10$ with smearing parameter\footnote{For details of the
  smearing we refer to subsection \ref{sec:potential_simulation_details}.}
$\lambda_s=0.2$ and subsequently projected into the $A_1^{++}$ channel.

In the two large simulations at $\beta=3.15$ and 3.40 we measure all
22 Wilson loop shapes up to length eight (see \cite{Berg:1983kp}) on
the same smearing levels mentioned before and project them into all 20
irreducible glueball channels.

A considerable part of the simulation time is used to measure all the
22 loop shapes.
Some of them may turn out to be superfluous in the sense that
they give a much worse signal/noise ratio than the others.
On the other hand, one is interested in having a set of operators as
large as possible to build up the wave function of the lowest glueball
state (more precisely, to cancel the unwanted contributions
from the neighbouring states in the spectrum).
Having measured all these operators will allow us to identify the
important loop shapes to be used in future simulations.
  
The projections of the loop shapes into the irreducible representations of the
cubic group are done according to the descriptions in
\cite{Berg:1983kp,cornwell:1984}.  The correlation matrix elements are then
constructed from the projected operators and Monte Carlo estimates are
obtained by averaging the measurements in each bin.  We measure all possible
polarizations\footnote{In analogy to choosing different magnetic quantum
  numbers $m$ for given angular momentum $l$ in the O(3) group.} in a given
channel and add them together in the correlation matrix. This eventually
suppresses the statistical noise more than just increasing the statistics
since the different polarizations are anti-correlated.

For the extraction of the glueball masses in the $A_1^{++}$
representation (which has the same quantum numbers as the vacuum) one
has to consider vacuum-sub\-tracted operators. For this purpose we
also measure the expectation values of all the operators.

\subsection{Details of the analysis}\label{sec:glueballs_analysis_details}
The glueball masses are extracted using the variational techniques 
described in appendix \ref{sec:var_tech}. Let us put some remarks which 
are related to the analysis of the glueball masses in particular.

As we are measuring a large number of operators (up to 145), normally some of
them contain large statistical noise.  Therefore we only keep a set of well
measured operators, on which the whole procedure is numerically stable and well
defined.

Another remark concerns the vacuum subtraction necessary in the
$A_1^{++}$ channel. To obtain vacuum-subtracted operators one usually
considers \linebreak 
$\phi^{\text{sub}}(\tau)=\phi(\tau)-\langle 0|\phi(\tau)|0\rangle$.
However, we follow a different strategy and treat the vacuum 
on the same footing as the other states in the vacuum channel.
As it turns out, the vacuum state can be separated in this procedure
with very high accuracy and it is safe to consider only the operator 
basis orthogonal to the vacuum in the fitting procedure. 
For this purpose we cut out the vacuum state obtained from solving 
the generalized eigenvalue equation
(\ref{eq:gen_eigenvalue_problem}), i.e.~we only consider the
correlation matrix\footnote{See appendix \ref{sec:var_tech} 
for notations.}
\begin{equation}
 C^K_{ij}(t) = (v_i, C^M(t) v_j), 
\end{equation}
with $i,j$ running from $i,j=2,\ldots, K \leq M$ in the further
analysis ($i=1$ being the vacuum state).
In our experience this strategy yields the most stable
subtraction of the vacuum contribution with respect to the statistical
fluctuations of the subtracted operators.

In the last step for extracting the glueball masses the large correlation
matrix is truncated to a $1 \times 1$ or $2 \times 2$ matrix, which is
subsequently fitted in the fit range $t_{\text{min}} \ldots t_{\text{max}}$
taking both temporal correlations and correlations among the operators into
account. The corresponding covariance matrix is calculated from jackknife
samples and the error is estimated using a jackknife procedure.  The choice of
$t_{\text{max}}$ is not crucial and is usually taken according to the relative
error of the matrix elements under consideration and the $\chi^2$-function.
More important is the correct choice of $t_{\text{min}}$. Since excited
glueball states are rather heavy we do not expect large contamination of the
ground state correlators from excited states even on time slice $t=1$ and
therefore $t_{\text{min}}=1$ is usually chosen.  In particular this choice is
safe if we fix $t_0=1$ and $t_1=2$ rather than $t_0=0$ and $t_1=1$ in the
variational method, eq.~(\ref{eq:gen_eigenvalue_problem}).  Indeed, in the
former case the $\chi^2$-function remains more stable when we increase
$t_{\text{min}}=1$ to $t_{\text{min}}=2$ as a check for the consistency of the
resulting masses (as an example take the results in table
\ref{tab:b286_fit_results}).

\subsection{Results}\label{sec:glueball_results}
The results of the fits to the glueball correlators are collected in
the appendix in tables \ref{tab:b286_fit_results} -- 
\ref{tab:large_b340_fit_results}.
We include the results
of different fitting ranges in $t$ in the tables in order to give an
impression of the stability of the fits.  In each channel the result
highlighted in boldface is our final choice and represents a most
reasonable mass for the given channel. These final mass estimates in
units of the lattice spacing are collected in table
\ref{tab:final_mass_estimates}.

\begin{figure}[htbp]
  \begin{center}
    \includegraphics[width=9cm,angle=-90]{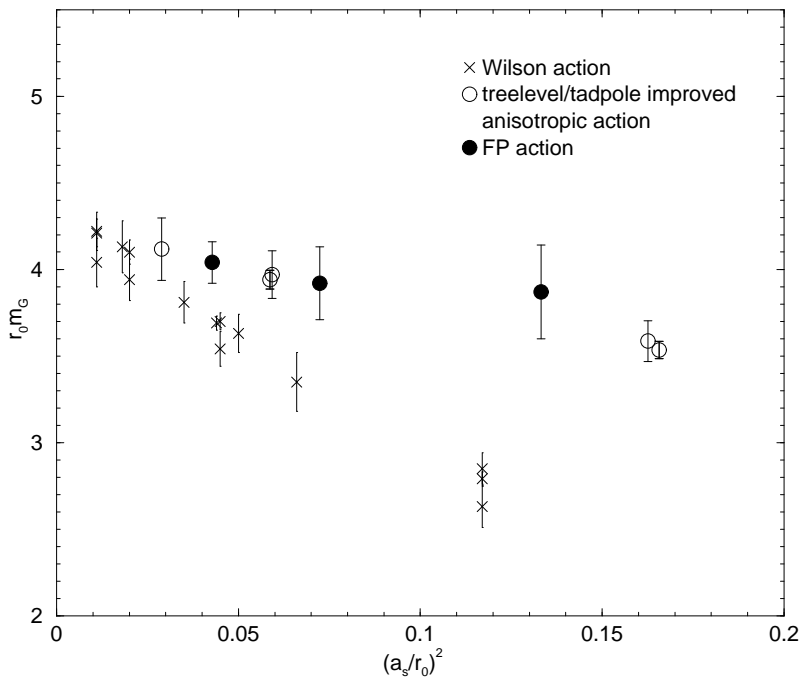}
    \caption{{}Glueball mass estimates for the $A_1^{++}$ channel. 
      Results from simulations of the Wilson action (crosses) and a 
      tree level/tadpole improved anisotropic action (empty circles) 
      are shown together with the results obtained with the FP action 
      (filled circles).}
    \label{fig:A1pp_final_mass}
  \end{center}
\end{figure}

\begin{figure}[htbp]
  \begin{center}
    \includegraphics[width=9cm,angle=-90]{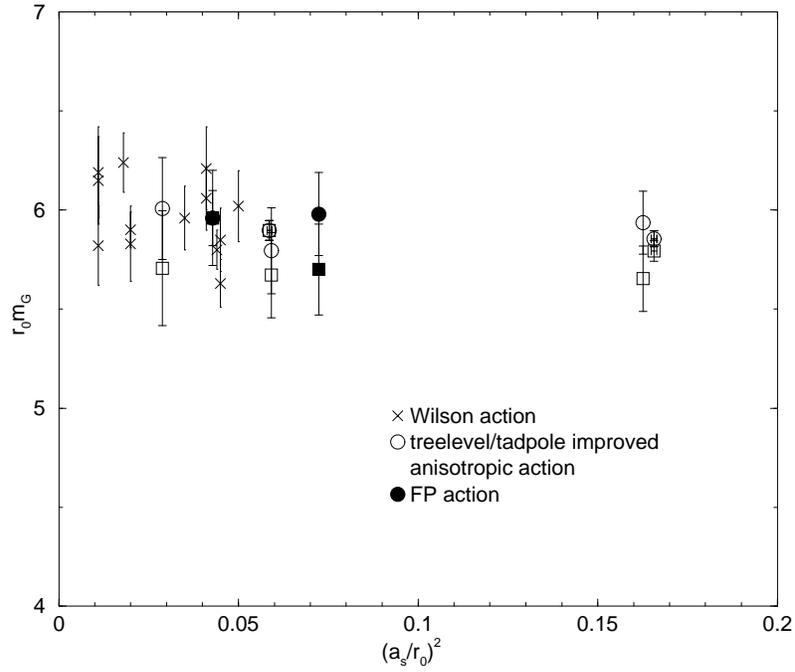}
    \caption{{}Glueball mass estimates for the $2^{++}$
    channel. Results from simulations of the Wilson action (crosses)
    and a tree level/tadpole improved anisotropic action (empty
    symbols) are shown together with the results obtained with the FP
    action (filled symbols). Squares and circles denote the $E^{++}$
    and $T_2^{++}$ mass estimates, respectively.}
    \label{fig:Epp_final_mass}
  \end{center}
\end{figure}

To compare the values it is convenient to use $r_0$ to set the
scale. In table \ref{tab:final_r0_mass_estimates} we list our
estimates of the glueball masses expressed in terms of $r_0$, while
figure \ref{fig:A1pp_final_mass} and \ref{fig:Epp_final_mass} show our
values for the $A_1^{++}$ and the $E^{++}$, $T_2^{++}$ channels,
respectively, together with results from different calculations with
the Wilson action (crosses)
\cite{Teper:1998kw,Bali:1993fb,Vaccarino:1999ku} and the calculations
of Morningstar and Peardon \cite{Morningstar:1997ff,Morningstar:1999rf} and
Liu \cite{Liu:2000ce} with a tree level/tadpole improved anisotropic
action (empty symbols).

To compare our results to the continuum values of the various
collaborations we resort to \cite{Wittig:1999kb} where the above
Wilson action results have been expressed in or converted to units of
$r_0$ using the interpolating formula for the Wilson action
\cite{Guagnelli:1998ud} and, whenever necessary, the continuum
extrapolation has been redone. Our continuum result for the $0^{++}$ glueball
mass is an extrapolation to the continuum using a fit function linear
in $(a/r_0)^2$.  The data in the other channels does not allow to do
an extrapolation, thus we simply quote the masses
obtained on the finest lattice ($a=0.10$~fm) in brackets. The
comparison of our results to the continuum values of the other groups
is listed in tables \ref{tab:comp1_gb} and \ref{tab:comp2_gb}.

Note that one observes  restoration
of the degeneracy within the statistical errors for the $2^{++}$ state as well
as for the $2^{-+}$ state. All our mass estimates agree with the best earlier
results within the statistical errors.

\begin{table}[htbp]
  \begin{center}
    \begin{tabular}{|llll|}
      \hline
      Collab. & $r_0 m_{0^{++}}$ & $r_0 m_{2^{++}}$  & year    \\
      \hline
      UKQCD \cite{Bali:1993fb}        & 4.05(16)    & 5.84(18)   & 1993 \\
      Teper \cite{Teper:1998kw}       & 4.35(11)    & 6.18(21)   & 1998 \\
      GF11 \cite{Vaccarino:1999ku}    & 4.33(10)    & 6.04(18)   & 1999 \\
      M\&P \cite{Morningstar:1999rf}  & 4.21(11)(4) & 5.85(2)(6) & 1999 \\
      Liu  \cite{Liu:2000ce}          & 4.23(22)    & 5.85(23)   & 2000 \\
      \hline
      FP action                       & 4.12(21)    &[5.96(24)]  & 2000 \\
      \hline
    \end{tabular}
    \caption{{}Comparison of the two lowest glueball masses in units
    of $r_0$. Our $2^{++}$ value is not extrapolated to the continuum
    but is the mass obtained at a lattice spacing $a=0.10$ fm.}
    \label{tab:comp1_gb}
  \end{center}
\end{table}
\begin{table}[htbp]
  \begin{center}
    \begin{tabular}{|lllll|}
      \hline
    Collab. & $r_0 m_{0^{-+}}$ & $r_0 m_{2^{-+}}$ & $r_0 m_{1^{+-}}$ & year \\
      \hline
    Teper \cite{Teper:1998kw}     & 5.94(68) & 8.42(78) & 7.84(62) & 1998 \\
    M\&P \cite{Morningstar:1999rf}& 6.33(7)(6) & 7.55(3)(8) & 7.18(4)(7) 
                                                                   & 1999 \\
    \hline
    FP action                     &[6.74(42)]&[8.00(35)]&[7.93(78)]& 2000 \\
    \hline
    \end{tabular}
    \caption{{}Comparison of glueball masses in units of $r_0$. 
      Values in brackets denote masses obtained at a lattice spacing
      $a=0.10$ fm and are not extrapolated to the continuum.}
    \label{tab:comp2_gb}
  \end{center}
\end{table}

Finally we convert the scalar and tensor glueball masses into physical units
using $r_0 \approx 0.49\, \text{fm} = (395 \text{MeV})^{-1}$.  We obtain 1627(83)
MeV for the $0^{++}$ and 2354(95) MeV for the $2^{++}$ glueball mass,
respectively. Note that the latter value corresponds to the glueball mass
measured at a lattice spacing $a=0.10$ fm.

As mentioned in the introduction, it is well known that glueball
masses are difficult to measure on the lattice. Indeed, we can barely
resolve higher lying glueball states and measuring excited states
becomes impossible at the lattice spacings currently available to
us. In this sense we can not really take advantage of the parametrized
FP action, which is intended to be used on coarse lattices.

One way around this difficulty is the use of anisotropic lattices,
where the lattice spacing in temporal direction is much smaller than
in spatial direction, $a_\tau \ll a_\sigma$.  The work on the application
of the FP approach to anisotropic lattice gauge actions is in
progress \cite{rufenacht:2000}.

\section{Summary}

In this work we have presented a new parametrization of the FP action of a
specific RGT. It uses simple plaquettes built from single
gauge links as well as from smeared (``fat'') links. It reproduces the
classical properties of the action excellently and respects approximate scale
invariance of instanton solutions. Since in addition to the FP action values
we parametrize the derivatives with respect to the gauge fields, local changes
of the action in a MC simulation are better re\-pre\-sen\-ted.

The parametrization has been optimized at lattice spacings suitable for
performing simulations on coarse lattices up to $a\approx 0.3$~fm.

For subjecting the action to scaling tests we have
determined its critical couplings $\beta_c$ on lattices with temporal
extensions $N_\tau=$ 2, 3 and 4. For each $N_\tau$ we have performed
simulations on several lattices for a finite size scaling study.
Furthermore, we have measured the static quark--antiquark potential at various
values of the gauge coupling corresponding to $a\approx 0.1-0.3$~fm. From the
potential we have extracted the commonly used reference scale $r_0$ and the
effective string tension $\sigma$ in order to check the scaling
behaviour of the parametrized FP action by means of the dimensionless
quantities $r_0 T_c$, $T_c/\sqrt{\sigma}$ and $r_0 \sqrt{\sigma}$. In all the
quantities we observe excellent scaling within the statistical errors, even
on our coarsest lattices.

Additionally, we have measured the glueball spectrum in all symmetry channels.
The $A_1^{++}$ channel, which shows particularly large lattice artifacts in
measurements with the Wilson gauge action, is an excellent candidate for
testing the improvements achieved with the parametrized FP action.
We observe scaling of the glueball masses and restoration of the rotational symmetry in
the $2^{++}$ and $2^{-+}$ channel within the statistical errors. For the
glueball masses we obtain 1627(83) MeV for
the $0^{++}$ glueball in the continuum and 2354(95) MeV for the $2^{++}$
glueball at a lattice spacing of $a=0.10$ fm.
\newline

\noindent {\bf Acknowledgements}
We would like to thank Peter Hasenfratz for useful suggestions and 
discussions. 

\pagebreak

\begin{appendix}
\setcounter{equation}{0}
\renewcommand{\theequation}{\mbox{A.\arabic{equation}}}
\setcounter{table}{0}
\renewcommand{\thetable}{\mbox{A.\arabic{table}}}

\section{The parametrization}

\subsection{Details of the parametrization}\label{app:parametrization}

Let us introduce the notation $S_\mu^{(\nu)}(n)$ for the sum of two staples of
gauge links in direction $\mu$ in the $\mu\nu$-plane:
\begin{multline}
  S_\mu^{(\nu)}(n)=
  U_\nu(n)U_\mu(n+\hat{\nu})U_\nu^\dagger(n+\hat{\mu}) \\
  +U_\nu^\dagger(n-\hat{\nu})U_\mu(n-\hat{\nu})U_\nu(n-\hat{\nu}+\hat{\mu})\,.
\end{multline}

Besides the usual symmetric smearing, we shall also use a non-symmetric
smearing. For the symmetric smearing define
\begin{equation}
Q_\mu^{\rm s}(n)=
\frac{1}{6}\sum_{\lambda\ne\mu}S_\mu^{(\lambda)}(n)-U_\mu(n) \,
\end{equation}
and\footnote{{}Note that $x_\mu(n)$ is negative: $-4.5 \le x_\mu(n) \le 0$.}
\begin{equation}
 x_\mu(n)=
{\rm Re\,Tr}\left( Q_\mu^{\rm s}(n)U_\mu^\dagger(n) \right) .
\end{equation}

To build a plaquette in the $\mu\nu$-plane from smeared links 
we introduce asymmetrically smeared links.  
First define\footnote{The argument $n$ is suppressed in the following.}
\begin{equation}
Q_\mu^{(\nu)}=
\frac{1}{4}\left( \sum_{\lambda\ne\mu,\nu}S_\mu^{(\lambda)}
+\eta(x_\mu)S_\mu^{(\nu)} \right) 
-\left( 1+\frac{1}{2}\eta(x_\mu) \right) U_\mu \,.
\end{equation}
Using these matrices we build the asymmetrically smeared links
\begin{equation}\label{eq:smeared_link_W}
W_\mu^{(\nu)}=
U_\mu+c_1(x_\mu)Q_\mu^{(\nu)}+
c_2(x_\mu)Q_\mu^{(\nu)}U_\mu^\dagger Q_\mu^{(\nu)} +\ldots \,.
\end{equation}
Here $\eta(x)$, $c_i(x)$ are polynomials:
\begin{equation}
\eta(x_\mu)=\eta^{(0)} + \eta^{(1)} x_\mu +\eta^{(2)} x_\mu^2 +\ldots
\end{equation}
and 
\begin{equation}
c_i(x_\mu)=c_i^{(0)} + c_i^{(1)} x_\mu +c_i^{(2)} x_\mu^2 +\ldots .
\end{equation}

\subsection{The ${\cal O}(a^2)$ Symanzik conditions}\label{app:Symanzik}
In this appendix we derive the ${\cal O}(a^2)$ Symanzik conditions
\cite{Symanzik:1983dc,Symanzik:1983gh,Weisz:1983zw,Weisz:1984bn,
Luscher:1985xn,Luscher:1985zq}
by considering constant non-Abelian gauge potentials. 
The formulas apply to general SU($N$).

In the continuum we have one scalar, gauge-invariant dimension-4 operator
\begin{align}
R_0 &= 
-\frac{1}{2}\sum_{\mu\nu} {\rm Tr}\left( {\cal F}_{\mu\nu}^2 \right),
\label{R0} \\
\intertext{and three dimension-6 operators:}
R_1 &= 
\frac{1}{2}\sum_{\mu\nu} \text{Tr}
\left( \left( D_\mu {\cal F}_{\mu\nu} \right)^2 \right), \label{R1}\\
R_2 &= 
\frac{1}{2}\sum_{\mu\nu\lambda} \text{Tr}
\left( \left( D_\mu {\cal F}_{\nu\lambda} \right)^2 \right), \label{R2}\\
R_3 &= 
\frac{1}{2}\sum_{\mu\nu\lambda} \text{Tr}
\left( D_\mu {\cal F}_{\mu\lambda} 
       D_\nu {\cal F}_{\nu\lambda}  \right).
\label{R3}
\end{align}

According to Symanzik \cite{Symanzik:1983dc,Symanzik:1983gh}
the ${\cal O}(a^2)$ lattice artifacts of an action are described by an
effective continuum action where to the usual continuum action
($\propto R_0$) additional terms proportional to
$a^2 R_1$, $a^2 R_2$ and $a^2 R_3$ are added with appropriate
coefficients.
In fact, the equations of motion are 
$\sum_\mu D_\mu {\cal F}_{\mu\lambda}=0$ hence the term with $R_3$
can be eliminated by a change of variables, hence does not affect
the lattice artifacts in on-shell quantities, e.g.~masses.
(Note that the static $q\bar{q}$ potential is an off-shell quantity,
it depends on the choice of the operators. For such quantities
one has to improve the operators as well to get rid of artifacts.)
The ${\cal O}(a^2)$ lattice artifacts of on-shell quantities can be 
eliminated in all orders of perturbation theory by adding to 
the original lattice action two additional terms which in the naive 
continuum limit are proportional to $a^2 R_1$ and $a^2 R_2$,  
with appropriate coefficients.
On the tree level the absence of ${\cal O}(a^2)$ artifacts means
that when one expands the lattice action in powers of $a$, for smooth
fields the coefficients of $R_1$ and $R_2$ vanish.
The coefficient of $R_3$ is not required to vanish (and usually
it does not for the FP action).

For the specific lattice gauge action ansatz considered in section
\ref{sec:parametrization} one 
obtains\footnote{{}From the non-linear parameters only the zeroth order
  coefficients contribute to the normalization and the ${\cal O}(a^2)$ 
  Symanzik condition. To keep notation simple we substitute 
  $c_i^{(0)}\rightarrow c_i$  and $\eta^{(0)} \rightarrow \eta$ in 
  the rest of this subsection.} 
\begin{multline}
\sum_{\mu<\nu} w_{\mu\nu} = \frac{1}{4} R_0(1+(4+2\eta)c_1)\\
+\frac{1}{12}R_1\left( 1-2c_1(1-4\eta)+
\frac{3}{2}(1-\eta)^2(c_1^{2}-2c_2) \right) \\
+\frac{1}{2}R_3\left( c_1+
\frac{1}{4}(1+2\eta)(c_1^{ 2}-2c_2) \right) \,.
\end{multline} 
The normalization condition is obtained from the coefficient of $R_0$,
\begin{equation}
 \label{eq:norm}
 p_{10} + p_{01} (1+(4+2\eta)c_1) =1 \,.
\end{equation}
The first ${\cal O}(a^2)$ Symanzik condition requires the coefficient
of $R_1$ to vanish,
\begin{equation}
  \label{eq:sym1}
p_{10} + p_{01} \left( 1-2c_1(1-4\eta)+
\frac{3}{2}(1-\eta)^2(c_1^{ 2}-2c_2) \right) = 0 \,.
\end{equation}
It is interesting to see that the operator $R_2$ is absent and hence the
second ${\cal O}(a^2)$ Symanzik condition is satisfied automatically for the
general ansatz considered here. 
(Note that when the FP action is expressed in terms of simple loops 
some of them give a nonzero contribution to $R_2$!)

\subsection{The quadratic approximation}\label{sec:quadratic_approximation}

The couplings of the FP action can be calculated analytically in the quadratic
approximation \cite{DeGrand:1995jk,Blatter:1996ti}. By fitting the leading
order nonlinear parameters $\eta^{(0)}$, $c_1^{(0)}, c_2^{(0)}$ and $p_{10}$,
$p_{01}$ to the quadratic part of the FP action we can check the flexibility
and the quality of the parametrization. Although the true FP action fulfills
the tree-level Symanzik conditions to all orders in $a$, an approximate
parametrization introduces small violations of all these conditions.  
However, one can exploit the freedom in the parametrization to correct for
this and to fulfill explicitly the ${\cal O}(a^2)$ on-shell Symanzik
conditions.  The linear parameters $p_{10}$ and $p_{01}$ are determined as
functions of $\eta^{(0)}$, $c_1^{(0)}$ and $c_2^{(0)}$ by 
eqs.~(\ref{eq:norm}) and (\ref{eq:sym1}).  
The fit to the exactly known quadratic approximation of the FP action 
yields the following result for the three nonlinear parameters:
\begin{equation}
  \label{rgt3}
  \eta^{(0)}=0.082 \,,\qquad c_1^{(0)}=0.282 \,, 
  \qquad c_2^{(0)}=0.054 \,,
\end{equation}
with the corresponding plaquette coefficients 
\begin{equation}
  \label{plaq}
  p_{10} = -0.368095 \,, \qquad
  p_{01} =  0.629227 \,.
\end{equation}
This action is denoted by ${\cal A}_0$ and is a good approximation 
to the FP action for sufficiently smooth fields.

\subsection{The parametrized FP action}\label{app:param_FP}
The following table collects the numerical values of the non-linear and linear
parameters describing the approximate FP action in the range of lattice
spacing $0.03\text{ fm} \lesssim a \lesssim 0.3\text{ fm}$.

The set of parameters consists of four non-linear parameters
$\eta^{(0)},c_1^{(0)},c_2^{(0)},c_3^{(0)}$ describing the asymmetrically
smeared links $W_\mu^{(\nu)}$ and fourteen linear parameters $p_{kl}$ with $0
< k+l \leq 4$. This set approximates reasonably well the true FP action in the
range of $a$ given above.  (For smaller fluctuations -- in the intermediate
steps of the parametrization -- we used polynomials for $\eta(x)$ and $c_i(x)$
up to forth order.)

The optimal non-linear parameters are found to be 
\begin{equation}
\eta^{(0)}=-0.038445 \,, \quad c_1^{(0)}=0.290643  \,, \quad
c_2^{(0)}=-0.201505 \,, \quad c_3^{(0)}=0.084679\,.
\nonumber
\end{equation}

The linear parameters are collected in table \ref{tab:ape_431_action}.

\begin{table}[htbp]
  \begin{center}
    \begin{tabular}{|c|rrrrr|}
\hline

     & $l=0$ & $l=1$  & $l=2$ & $l=3$ & $l=4$ \\
\hline
 $k=0$  &            &  0.442827 & 0.628828 & -0.677790 &  0.176159 \\
 $k=1$  &  0.051944  & -0.918625 & 1.064711 & -0.275300 &           \\
 $k=2$  &  0.864881  & -0.614357 & 0.165320 &           &           \\
 $k=3$  & -0.094366  & -0.020693 &          &           &           \\
 $k=4$  &  0.022283  &           &          &           &           \\
\hline
  \end{tabular}
    \caption{{}Linear parameters $p_{kl}$ of the parametrized FP action.}
    \label{tab:ape_431_action}
  \end{center}
\end{table}

\setcounter{equation}{0}
\renewcommand{\theequation}{\mbox{B.\arabic{equation}}}
\setcounter{table}{0}
\renewcommand{\thetable}{\mbox{B.\arabic{table}}}

\section{Variational techniques} \label{sec:var_tech} 
In a Monte Carlo simulation we measure the $N \times N$
correlation matrix
\begin{equation}
  \label{eq:full_correlation}
  C_{\alpha \beta}(t) =  \langle 0| {\cal O}_\alpha(t)
  {\cal O}^\dagger_\beta(0) |0 \rangle \,.
\end{equation}
To determine the coefficients $v_\alpha$ of the linear combination 
$\sum_{\alpha=1}^N v_\alpha {\cal O}_\alpha$ which has the largest 
overlap to the ground state relative to the excited states 
one has to minimize the effective mass given by
\begin{equation}
  \label{eq:minimizing_eff_mass}
  m(t_0,t_1) = 
-\ln\left[ \frac{(v, C(t_1)v)}{(v, C(t_0)v)}\right]/(t_1-t_0).
\end{equation}
The vector $v$ is obtained by solving the generalized eigenvalue equation
\cite{Michael:1985ne,Luscher:1990ck}
\begin{equation}
  \label{eq:gen_eigenvalue_problem}
  C(t_1) v = \lambda(t_0,t_1) C(t_0) v \,,
\end{equation}
where $0 \le t_0 < t_1$.

Assume first that only the lowest lying $N$ states contribute to
$C(t)$, i.e.
\begin{equation}
\label{eq:C_N}
C_{\alpha \beta}(t)=\sum_{n=1}^{N} \text{e}^{-E_nt}\psi_{n\alpha}
\psi_{n\beta}^* \,,
\end{equation}
where $E_1 \le E_2 \le \ldots \le E_N$ are the energy levels in the
given symmetry channel and 
$\psi_{n\alpha}=\langle 0 | O_\alpha | n \rangle$ is the 
``wave function'' of the corresponding state.
The solution of eq.~(\ref{eq:gen_eigenvalue_problem}) is given
by the set of vectors $\{ v_n \}$ dual to the wave functions,
i.e.~$(v_n,\psi_m)=\delta_{nm}$.
Multiplying eq.~(\ref{eq:C_N}) by $v_n$ one obtains
\begin{equation}
\label{eq:Cv}
C(t)v_n=\text{e}^{-E_nt}\psi_n = 
\text{e}^{-E_n(t-t_0)}\text{e}^{-E_nt_0}\psi_n=
\text{e}^{-E_n(t-t_0)}C(t_0)v_n \,.
\end{equation}
This gives $\lambda_n(t_0,t_1)=\exp( -E_n(t_1-t_0) )$ for the eigenvalues
in eq.~(\ref{eq:gen_eigenvalue_problem}).
Of course, contributions from states with $n>N$ and statistical
fluctuations distort eq.~(\ref{eq:C_N}), therefore the stability
of eq.~(\ref{eq:gen_eigenvalue_problem}) is an important issue.

Observe that eq.~(\ref{eq:gen_eigenvalue_problem}) is well defined only 
for positive definite $C(t_0)$. 
Because of statistical fluctuations, however, the measured correlation 
matrix $C(t_0)$ is not necessarily positive for $t_0>0$.
This is the reason why one usually considers only the $t_0=0$
case in applying the variational method, especially with a large
number of operators.
On the other hand, it is obvious that $C(0)$ is contaminated by highly 
excited states and contains only restricted information on the low lying 
part of the spectrum. Therefore it is desirable to take $t_0>0$.
This can be achieved in the following way \cite{Balog:1999ww}.

We first diagonalize $C(t_0)$,
\begin{equation}
  \label{eq:C0_diagonalization}
  C(t_0) \varphi_i  = \lambda_i \varphi_i, \quad \lambda_1 \geq \ldots \geq
  \lambda_N,
\end{equation}
and project the correlation matrices to the space of eigenvectors
corresponding to the $M$ highest eigenvalues,
\begin{equation}
  \label{eq:first_truncation}
  C^M_{ij}(t) = (\varphi_i, C(t) \varphi_j), \quad  i,j=1,\ldots, M.
\end{equation}
By choosing the operator space too large we introduce numerical instabilities
caused by very small (even negative) eigenvalues with large statistical errors
due to the fact that the chosen operator basis is not 
sufficiently independent on the given MC sample. 
By choosing $M$ appropriately we can get rid of those unstable
modes while still keeping all the physical information. In this way we 
render the generalized eigenvalue problem well defined.

Of course the final result should not depend on the choice of $M$ and one has
to take care in each case that this is really the case. Our observation is that for
any acceptable statistics one always finds a plateau in $M$ for which the
extracted masses are stable under variation of $M$.

In a next step we determine the vectors $v_n$, $n=1, \ldots, M$ through the
generalized eigenvalue equation in the truncated basis:
\begin{equation}
\label{genev_CM}
  C^M(t_1) v_n = e^{-E_n(t_1-t_0)} C^M(t_0) v_n \,.
\end{equation}
This equation yields the spectrum $E_n$. However, the procedure
-- although it is exact for a correlation matrix which
has {\em exactly} the form in eq.~(\ref{eq:C_N}) --
is highly non-linear, and a small statistical fluctuation can be 
enhanced by it and cause a systematic shift in the energy values 
obtained, even when the instabilities are avoided by the truncation
to $M<N$.

In order to avoid this pitfall we use the (approximate) dual
vectors $v_n$ obtained from eq.~(\ref{genev_CM}) to restrict
the problem to a smaller, therefore more stable subspace.

Define the new correlation matrix of size $K\times K$ (with $K\le M$) by
\begin{equation}
  \label{eq:second_truncation}
 C^K_{ij}(t) = (v_i, C^M(t) v_j), \quad i,j=1,\ldots, K \leq M \,.
\end{equation}
The steps performed until now can be thought of as a preparation for 
choosing the appropriate set of operators, i.e.~linear combinations
of original ${\cal O}_\alpha$ operators which effectively
eliminate the higher states. The correlation matrix
$C^K_{ij}(t)$ is then considered as a primary, unbiased object.

The next step is to fit $C^K_{ij}(t)$ in the range 
$t = t_{\text{min}} \ldots t_{\text{max}}$ using the ansatz
\begin{equation}\label{eq:ansaetze}
  \widetilde{C^K_{ij}}(t;\{\psi, E\}) = \sum_{n=1}^{K} e^{-E_n t} \psi_{ni} \psi_{nj}^* \,,
\end{equation}
where $\psi_{ni}$, $E_n$ are the free parameters to be fitted.

Usually we choose $K=1$  and $2$. For the $A_1^{++}$ glueball, however,
$K=2$ and $3$ are chosen since we do not subtract the vacuum contribution 
$\langle {\cal O}_\alpha \rangle \langle {\cal O}_\beta \rangle^*$ 
from the correlators but consider instead  the vacuum state
together with the glueball states in this channel
(cf.~remarks in section \ref{sec:glueballs_analysis_details}).

In the fitting step we use a correlated $\chi^2$ fit
which takes into account the correlation between $C^K_{ij}(t)$
and $C^K_{i'j'}(t')$, i.e.~using the inverse of the corresponding 
covariance matrix  $\text{Cov}(i,j,t; i',j',t')$ as a weight 
in the definition of $\chi^2$.
This has the advantage over the usual (uncorrelated) $\chi^2$
that the value of the latter can be artificially small
if the quantities to be fitted are strongly correlated.
Note however, that (as usually with sophisticated methods)
the correlated $\chi^2$ fit can have its own instabilities
if the number of data is not sufficiently large
\cite{Michael:1994yj,Michael:1995br}.

\setcounter{equation}{0}
\renewcommand{\theequation}{\mbox{C.\arabic{equation}}}
\setcounter{table}{0}
\renewcommand{\thetable}{\mbox{C.\arabic{table}}}

\section{Simulation parameters and results}
\subsection{Deconfining phase transition}

\begin{table}[htbp]
  \begin{center}
    \begin{tabular}{ccccr}
\hline
\hline
lattice size & $\beta$ & sweeps & $\tau_p$ & $\tau_{\text{int}}$ \\
\hline
$2 \times 10^3$ & 2.3550 & 30000 &      & 260 \\
                & 2.3575 & 30000 & 4300 & 283 \\
                & 2.3560 & 30000 & 4600 & 280 \\
\hline                                       
$2 \times 8^3$  & 2.3300 & 14240 &      &  29 \\
                & 2.3500 & 10144 &      &  93 \\
                & 2.3550 &  5120 &      & 127 \\
                & 2.3575 & 12288 & 1400 & 202 \\
                & 2.3700 & 10144 &      & 114 \\ 
\hline                                       
$2 \times 6^3$  & 2.3250 &  8096 &      &  35 \\
                & 2.3500 & 14144 &  650 & 105 \\
                & 2.3600 & 10000 &  700 &  96 \\
                & 2.3750 & 10144 &      &  39 \\
\hline
\hline
\end{tabular}
    \caption{{}Run parameters of the critical temperature simulations at
      $N_\tau=$~2 including the persistence time $\tau_p$ and the integrated
      autocorrelation time $\tau_{\text{int}}$ of the Polyakov loop operator.}
    \label{tab:sim_details_N_t=2}
  \end{center}
\end{table}

\begin{table}[htbp]
  \begin{center}
    \begin{tabular}{ccccr}
\hline
\hline
lattice size & $\beta$ & sweeps & $\tau_p$ & $\tau_{\text{int}}$ \\
\hline
$3 \times 12^3$ & 2.675 & 25000 &      & 114 \\
                & 2.680 & 45000 & 3200 & 188 \\
                & 2.685 & 24000 &      &  96 \\
                & 2.690 & 20000 &      &  53 \\
\hline                                      
$3 \times 10^3$ & 2.670 & 18000 &      &  67 \\
                & 2.680 & 42000 & 2300 &  89 \\
                & 2.685 & 48000 & 2400 & 104 \\
                & 2.690 & 27000 &      &  85 \\
\hline                                      
$3 \times 8^3$  & 2.650 & 10096 &      &  43 \\
                & 2.660 & 10000 &      &  48 \\ 
                & 2.670 & 26000 &      &  41 \\ 
                & 2.680 & 30000 & 1400 &  64 \\
                & 2.690 & 19000 &      &  53 \\
                & 2.710 & 10000 &      &  35 \\
\hline
\hline
\end{tabular}
    \caption{{}Run parameters of the critical temperature simulations at
      $N_\tau=$~3 including the persistence time $\tau_p$ and the integrated
      autocorrelation time $\tau_{\text{int}}$ of the Polyakov loop operator.}
    \label{tab:sim_details_N_t=3}
  \end{center}
\end{table} 
\begin{table}[htbp]
  \begin{center}
    \begin{tabular}{ccccr}
\hline
\hline
lattice size & $\beta$ & sweeps & $\tau_p$ & $\tau_{\text{int}}$ \\
\hline
$4 \times 14^3$ & 2.917 & 50405 & 4300 & 62 \\
                & 2.922 & 51812 & 4700 & 67 \\
                & 2.930 & 44607 &      & 64 \\
\hline                                     
$4 \times 12^3$ & 2.850 & 15000 &      & 19 \\
                & 2.890 & 15000 &      & 30 \\
                & 2.910 & 33000 &      & 34 \\
                & 2.920 & 33000 & 3700 & 66 \\
                & 2.930 & 15000 &      & 38 \\
\hline                                     
$4 \times 10^3$ & 2.850 & 10000 &      & 22 \\
                & 2.880 & 16000 &      & 37 \\
                & 2.890 & 21124 &      & 18 \\
                & 2.900 & 35000 &      & 34 \\
                & 2.910 & 35000 & 2100 & 36 \\
                & 2.920 & 20000 &      & 39 \\
\hline
\hline
\end{tabular}
    \caption{{}Run parameters of the critical temperature simulations at
      $N_\tau=$~4 including the persistence time $\tau_p$ and the integrated
      autocorrelation time $\tau_{\text{int}}$ of the Polyakov loop operator.}
    \label{tab:sim_details_N_t=4}
  \end{center}
\end{table}
\clearpage

\subsection{Static quark-antiquark potential}
\begin{table}[htbp]
  \begin{center}
    \begin{tabular}{cccccc}
      \hline
      \hline
      $\beta$  & lattice volume  & lattice size [fm] & \# measurements \\
      \hline                   
      3.400    &  $14^4$  & 1.45 &  $43\times 90$ \\
      3.150    &  $12^4$  & 1.61 &  $42\times 50$ \\
      2.927    &  $14^4$  & 2.39 &  $40\times 40$ \\
      2.860    &  $10^4$  & 1.84 &  $43\times 90$ \\
      2.680    &  $12^4$  & 2.72 &  $51\times 40$ \\
      2.361    &  $12^4$  & 4.02 &  $57\times 40$ \\
      \hline
      \hline
    \end{tabular}
    \caption{{}Run parameters for the simulations of the static
      quark-antiquark potential.}
    \label{run_parameters_for_potentials}
  \end{center}
\end{table}
\begin{table}[htbp]
  \begin{center}
    \begin{tabular}{cccccc}
      \hline
      \hline
      $\beta$ & fit range & $a V_0$ & $\alpha$ & $\sigma a^2$ & $\chi^2/N_{\text{DF}}$\\
      \hline
      3.400 & 2 - 6 &  0.781(1)  & 0.251(9)  & 0.063(1) & 1.02 \\
      3.150 & 2 - 5 &  0.820(15) & 0.285(20) & 0.099(3) & 0.75 \\
      2.927 & 2 - 6 &  0.812(16) & 0.272(20) & 0.161(3) & 1.35 \\
      2.860 & 1 - 4 &  0.801(5)  & 0.262(3)  & 0.189(2) & 1.17 \\
      2.680 & 1 - 4 &  0.777(5)  & 0.255(4)  & 0.287(2) & 0.43 \\
      2.680 & 2 - 6 &  0.778(41) & 0.256(54) & 0.287(7) & 0.65 \\
      2.361 & 1 - 4 &  0.615(11) & 0.179(8)  & 0.629(4) & 0.99 \\
      2.361 & 2 - 5 &  0.59(11)  & 0.15(13)  & 0.634(22)& 1.41 \\ 
      \hline
      \hline
    \end{tabular}
    \caption{{}Results from global correlated fits of the form 
      (\ref{eq:potential_ansatz}) to the static quark potentials. 
      The second column indicates the fit range in $r$ and the last 
      column $\chi^2$ per degree of freedom, $\chi^2/N_{\text{DF}}$.}
    \label{tab:pot_fit_results}
  \end{center}
\end{table}

\begin{table}[htbp]
  \begin{center}
    \begin{tabular}{cccclc}
      \hline
      \hline
      $\beta$ & $r$ & $M$ & fit range & \phantom{     } $V(r)$ & $\chi^2/N_{\text{DF}}$ \\
      \hline
      3.400   & 1 & 5 & 2 - 6 & 0.5874(2) & 0.76\\
              & 2 & 5 & 2 - 6 & 0.7804(5) & 2.19\\
              & 3 & 5 & 3 - 6 & 0.885(2)  & 1.27\\
              & 4 & 3 & 3 - 6 & 0.969(3)  & 1.24\\
              & 5 & 4 & 2 - 6 & 1.046(4)  & 0.91\\
              & 6 & 4 & 2 - 5 & 1.116(8)  & 0.38\\
              & 7 & 3 & 3 - 6 & 1.17(2)   & 0.18\\
\hline
      3.150   & 1 & 5 & 3 - 5 & 0.6405(3) & 0.77\\
              & 2 & 4 & 2 - 6 & 0.8756(5) & 0.63\\
              & 3 & 4 & 2 - 6 & 1.023(1)  & 0.64\\
              & 4 & 3 & 2 - 5 & 1.147(2)  & 0.15\\
              & 5 & 3 & 2 - 6 & 1.258(3)  & 0.84\\
              & 6 & 3 & 2 - 6 & 1.38(1)   & 1.08\\
\hline
      2.927   & 1 & 4 & 2 - 7 & 0.7032(2) & 0.42 \\ 
              & 2 & 3 & 2 - 7 & 0.9969(5) & 0.65 \\
              & 3 & 3 & 2 - 7 & 1.202(2)  & 0.56 \\
              & 4 & 4 & 2 - 5 & 1.383(5)  & 0.31 \\
              & 5 & 3 & 2 - 7 & 1.560(8)  & 0.81 \\
              & 6 & 3 & 2 - 5 & 1.71(2)   & 0.82 \\
              & 7 & 2 & 2 - 6 & 1.92(3)   & 1.28 \\
\hline
      2.860   & 1 & 3 & 2 - 4 & 0.7267(4) & 1.50\\
              & 2 & 3 & 1 - 4 & 1.047(1)  & 0.56\\
              & 3 & 4 & 1 - 4 & 1.278(2)  & 0.68\\
              & 4 & 2 & 2 - 4 & 1.488(5)  & 0.30\\
              & 5 & 3 & 2 - 4 & 1.67(2)   & 0.68\\
\hline
      2.680   & 1 & 4 & 2 - 6 & 0.8091(3) & 0.21\\
              & 2 & 4 & 2 - 6 & 1.2231(9) & 0.98\\
              & 3 & 4 & 2 - 6 & 1.553(3)  & 0.33\\
              & 4 & 3 & 1 - 5 & 1.862(3)  & 0.33\\
              & 5 & 2 & 2 - 6 & 2.15(3)   & 0.89\\
              & 6 & 2 & 2 - 5 & 2.51(8)   & 0.14\\
\hline
      2.361   & 1 & 3 & 2 - 5 & 1.0641(6) & 0.33\\
              & 2 & 3 & 1 - 6 & 1.783(1)  & 0.31\\
              & 3 & 2 & 1 - 5 & 2.443(4)  & 0.75\\
              & 4 & 2 & 1 - 6 & 3.09(2)   & 0.84\\
              & 5 & 1 & 1 - 5 & 3.73(6)   & 2.31\\
              & 6 & 1 & 1 - 6 & 4.5(3)    & 0.44\\
      \hline
      \hline
    \end{tabular}
    \caption{{}Potential values extracted from fits of the form 
      $Z(r) \exp(-t V(r))$ to the ground state of the string correlators. 
      For each $\beta$-value and distance $r$ we list the plateau regions 
      (fit range $t_{\text{min}}-t_{\text{max}}$), the extracted potential 
      values $V(r)$ and the $\chi^2$ per degree of freedom, 
      $\chi^2/N_{\text{DF}}$.  Note that $t_0=1$ and $t_1=2$ was chosen in 
      all cases. The column entitled with $M$ denotes the number 
      of operators kept after the first truncation.}
    \label{tab:pot_values_all_betas}
  \end{center}
\end{table}

\clearpage

\subsection{Glueballs}
\begin{table}[htbp]
  \begin{center}
    \begin{tabular*}{\textwidth}{c@{\extracolsep{\fill}}cccccc}
      \hline \hline
      $\beta$ & lattice & $a[\text{fm}]$ & $r_0/a$ &  \# measurements\\
      \hline
      3.40 & $14^4$ & 0.10 & 4.83(4) & $206\times 70$ \\
      3.15 & $12^4$ & 0.13 & 3.72(2) & $202\times 50$ \\
      2.86 & $10^4$ & 0.18 & 2.74(1) & $160\times 50$ \\
      \hline \hline
    \end{tabular*}
    \caption{{}Run parameters of the glueball simulations. Values for the
      coupling $\beta$, the lattice size and the obtained statistics are
      listed. The estimate of the hadronic scale $r_0$ in terms of 
      the lattice spacing $a$ is given as well as the approximate lattice 
      spacing in fermi.}
    \label{tab:glueballs_sim_details}
  \end{center}
\end{table}
\begin{table}[htbp]
  \begin{center}
    \begin{tabular}{cccccl}
      \hline
      \hline
      Channel & $t_0/t_1$ & $M$ & fit range &
      $\chi^2/N_{\text{DF}}$ & energies \\
      \hline
      $A_1^{++}$ & 1/2 & 3  & 1 - 4 & 0.02 & \bf{1.41(10)} \\
                 &     &    & 2 - 4 & 0.02 & 1.40(38) \\
                 & 0/1 & 25 & 1 - 4 & 0.56 & 1.38(8) \\
                 &     &    & 2 - 4 & 0.36 & 1.50(40) \\
                 \hline
                 \hline
    \end{tabular}
    \caption{{}Results from fits to the $\beta=2.86$ glueball correlator on
      the $10^4$ lattice. Only five loop shapes were measured on 5 different
      smearing schemes.}
    \label{tab:b286_fit_results}
  \end{center}
\end{table}
\begin{table}[htbp]
  \begin{center}
    \begin{tabular}{cccccl}
      \hline
      \hline
      Channel & $t_0/t_1$ & $M$ & fit range &
      $\chi^2/N_{\text{DF}}$ & energies \\
      \hline
      $A_1^{++}$ & 1/2 & 5  & 1 - 3 & 0.61 & \bf{1.03(3)} \\
                 &     &    & 2 - 3 & 0.00 & 1.10(10) \\
                 &     &    & 1 - 4 & 2.02 & 1.03(3) \\
                 &     &    & 2 - 4 & 1.07 & \bf{1.12(11)} \\
                 & 0/1 & 25 & 1 - 4 & 1.62 & 1.02(3) \\
                 &     &    & 2 - 4 & 0.02 & 1.12(9) \\
      $E^{++}$   & 1/2 & 4  & 1 - 3 & 1.26 & \bf{1.53(6)} \\
                 & 0/1 & 48 & 1 - 3 & 1.41 & 1.46(5)\\ 
      $T_2^{++}$ & 1/2 & 4  & 1 - 3 & 0.68 & \bf{1.61(6)} \\
                 & 0/1 & 48 & 2 - 4 & 1.32 & 1.83(23) \\           
      $A_1^{-+}$ & 1/2 & 3  & 1 - 3 & 0.84 & \bf{1.65(18)} \\
      $E^{-+}$   & 1/2 & 3  & 1 - 3 & 0.00 & \bf{1.97(20)} \\
                 & 0/1 & 15 & 1 - 3 & 0.09 & 2.06(16) \\
      $T_2^{-+}$ & 1/2 & 5  & 1 - 3 & 0.00 & 1.39(27) \\
                 & 0/1 & 22 & 1 - 3 & 0.00 & \bf{1.92(11)} \\
      $T_1^{+-}$ & 1/2 & 4  & 1 - 3 & 2.70 & \bf{2.10(18)} \\
                 & 0/1 & 25 & 1 - 3 & 0.05 & \bf{2.04(12)} \\
                 \hline
                 \hline
    \end{tabular}
    \caption{{}Results from fits to the $\beta=3.15$ glueball correlators on
      the $12^4$ lattice: $t_0$/$t_1$ are used in the generalized eigenvalue
      problem, $M$ denotes the number of operators kept after 
      the truncation in $C(t_0)$.}
    \label{tab:b315_fit_results}
  \end{center}
\end{table}
\begin{table}[htbp]
  \begin{center}
    \begin{tabular}{cccccl}
      \hline
      \hline
      Channel & $t_0/t_1$ & $M$ & fit range &
      $\chi^2/N_{\text{DF}}$ & energies \\
      \hline
      $A_1^{++}$ & 1/2 & 6  & 1 - 4 & 0.79 & \bf{0.84(2)} \\
                 & 0/1 & 30 & 1 - 4 & 0.54 & 0.84(2) \\
      $E^{++}$   & 1/2 & 11 & 1 - 4 & 0.03 & \bf{1.23(5)} \\
                 &     &  8 & 1 - 4 & 0.19 & 1.27(3) \\
                 & 0/1 & 60 & 1 - 4 & 0.02 & 1.23(2) \\
      $T_2^{++}$ & 1/2 & 5  & 1 - 4 & 0.40 & \bf{1.23(3)} \\
                 &     & 7  & 1 - 4 & 0.16 & 1.20(3) \\
                 & 0/1 & 48 & 1 - 4 & 1.16 & 1.25(2) \\
      $A_1^{-+}$ & 1/2 & 3  & 1 - 3 & 0.24 & \bf{1.40(9)} \\
                 & 0/1 & 15 & 1 - 3 & 0.12 & 1.46(5) \\
                 &     & 15 & 2 - 4 & 0.10 & 1.38(20) \\
      $E^{-+}$   & 1/2 & 3  & 1 - 3 & 0.34 & \bf{1.68(7)} \\
      $T_2^{-+}$ & 1/2 & 4  & 1 - 3 & 0.09 & \bf{1.63(7)} \\
      $T_1^{+-}$ & 1/2 & 8  & 1 - 3 & 2.49 & \bf{1.64(16)} \\
                 &     & 6  & 1 - 3 & 0.17 & 1.76(10) \\
                 & 0/1 & 25 & 1 - 3 & 0.07 & 1.65(6) \\ 
                 \hline
                 \hline
    \end{tabular}
    \caption{{}Results from fits to the $\beta=3.40$ glueball correlators on
      the $14^4$ lattice: $t_0$/$t_1$ are used in the generalized eigenvalue
      problem, $M$ denotes the number of operators kept after the
      truncation in $C(t_0)$.} 
    \label{tab:large_b340_fit_results}
  \end{center}
\end{table}
\begin{table}[htbp]
  \begin{center}
    \begin{tabular}{clll}
    \hline 
    \hline
               &  $\beta=2.86$ & $\beta=3.15$ &  $\beta=3.40$ \\
    \hline                                                    
    $A_1^{++}$ &  1.41(10)    & 1.05(6)    &  0.84(2)    \\
    $E^{++}$   &               & 1.53(6)    &  1.23(5)    \\
    $T_2^{++}$ &               & 1.61(6)    &  1.23(3)    \\
    $A_1^{-+}$ &               & 1.65(18)     &  1.40(9)    \\
    $E^{-+}$   &               & 1.97(20)     &  1.68(7)    \\
    $T_2^{-+}$ &               & 1.92(11)     &  1.63(7)    \\
    $T_1^{+-}$ &               & 2.10(18)     &  1.64(16)     \\
    \hline 
    \hline
    \end{tabular}
    \caption{{}Final glueball mass estimates in terms of the lattice spacing,
    $a \,m_G$.} 
    \label{tab:final_mass_estimates}
  \end{center}
\end{table}
\begin{table}[htbp]
  \begin{center}
    \begin{tabular}{cclll}
    \hline 
    \hline
           & $J$ & $\beta=2.86$ & $\beta=3.15$     & $\beta=3.40$ \\
    \hline                                                         
    $A_1^{++}$ & 0 & 3.87(27)     &  3.92(23)    &  4.04(12)   \\
    $E^{++}$   & 2 &              &  5.70(23)    &  5.96(24)   \\
    $T_2^{++}$ & 2 &              &  5.98(21)    &  5.96(14)   \\
    $A_1^{-+}$ & 0 &              &  6.13(67)    &  6.74(42)   \\
    $E^{-+}$   & 2 &              &  7.32(74)    &  8.12(35)   \\
    $T_2^{-+}$ & 2 &              &  7.14(41)    &  7.88(35)   \\
    $T_1^{+-}$ & 1 &              &  7.81(67)    &  7.93(78)   \\
    \hline 
    \hline
    \end{tabular}
    \caption{{}Final glueball mass estimates in terms of $r_0$, $r_0 m_G$. 
      The continuum spin interpretation of each channel is denoted by $J$.}
    \label{tab:final_r0_mass_estimates}
  \end{center}
\end{table}

\clearpage
\end{appendix}

\end{document}